\documentclass[a4paper]{iopart}
\usepackage[table]{xcolor}
\usepackage{textcomp}
\usepackage{amsmath}
\usepackage{amssymb}
\usepackage{graphicx} 
\usepackage{pgf}
\usepackage{epstopdf}
\usepackage{iopams}

\usepackage[T1]{fontenc}
\usepackage[utf8x]{inputenc}
\inputencoding{utf8x}
\inputencoding{latin1}
\usepackage[output-decimal-marker={,}]{siunitx}
\usepackage{caption}
\usepackage{siunitx}  
\usepackage{bbold} 
\usepackage{mathtools}
\usepackage{subfigure}
\usepackage{wrapfig}
\usepackage{multirow}
\usepackage{microtype}
\usepackage{emptypage}
\usepackage{amsthm}
\usepackage{url}
\usepackage{pstricks}
\usepackage{amsfonts}
\usepackage{picture}

\usepackage{empheq} 
\usepackage{mparhack,fixltx2e,relsize}      
\usepackage{enumitem}  

\usepackage{float} 
                       
\usepackage{cite} 

\usepackage{titlesec}
\usepackage[titletoc,toc,title]{appendix} 

\graphicspath{{Figures/PNG/}{Figures/PDF/}{Figures/EPS/}{Figures/TEX/}{Figures/}} 

\definecolor{darkspringgreen}{rgb}{0.09, 0.45, 0.27}
\definecolor{islamicgreen}{rgb}{0.0, 0.56, 0.0}

\usepackage[colorlinks, hyperfootnotes = false, pdfpagelabels, colorlinks = true, linktocpage = true, pdfstartpage = 1, pdfstartview = FitV, breaklinks = true, pdfpagemode = UseNone, pageanchor= true, pdfpagemode = UseOutlines, hypertexnames=true,pdfhighlight=/0,   citecolor = islamicgreen,  linkcolor=blue, urlcolor=blue]{hyperref}

\usepackage{empheq} 

\usepackage{enumitem}

\def\be{\begin{equation}}
\def\ee{\end{equation}}

\def\bea{\begin{eqnarray}}
\def\eea{\end{eqnarray}}

\usepackage{etoolbox}

\makeatletter
\def\@mkboth#1#2{}
\newlength\appendixwidth
\preto\appendix{\addtocontents{toc}{\protect\patchl@section}}
\newcommand{\patchl@section}{%
  \settowidth{\appendixwidth}{\textbf{Appendix}}%
  \addtolength{\appendixwidth}{1.5em}%
  \patchcmd{\l@section}{1.5em}{\appendixwidth}{}{\ddt}%
}
\makeatother

\makeatletter
\def\@mkboth#1#2{}
\preto\appendix{\addtocontents{toc}{\protect\patchl@subsection}}
\newcommand{\patchl@subsection}{%
  \settowidth{\appendixwidth}{\textbf{Appendix}}%
  \addtolength{\appendixwidth}{10.5em}%
}
\makeatother

\def\ro{\hat{\rho}}

\def\bra#1{\langle #1 |}
\def\ket#1{| #1 \rangle}
\def\tr{\text{tr}}

\def\nn{\nonumber}

\def \Tr{\text{Tr}}

\def \sigmaup{
    \begin{pmatrix}
    1 & 0\\
    0 & 0
    \end{pmatrix}}
\def \sigmadown{
    \begin{pmatrix}
    0 & 0\\
    0 & 1
    \end{pmatrix}}
\def \sigmaplus{
    \begin{pmatrix}
    0 & 1\\
    0 & 0
    \end{pmatrix}}
\def \sigmaminus{
    \begin{pmatrix}
    0 & 0\\
    1 & 0
    \end{pmatrix}}    

\def \sigmaup{
    \begin{pmatrix}
    1 & 0\\
    0 & 0
    \end{pmatrix}}
\def \sigmadown{
    \begin{pmatrix}
    0 & 0\\
    0 & 1
    \end{pmatrix}}
\def \sigmaplus{
    \begin{pmatrix}
    0 & 1\\
    0 & 0
    \end{pmatrix}}
\def \sigmaminus{
    \begin{pmatrix}
    0 & 0\\
    1 & 0
    \end{pmatrix}}    

\def \rozero{
    \begin{pmatrix}
    a^2 & 0\\
    0 & b^2
    \end{pmatrix}}
\def \rouno{ \frac 12 
    \begin{pmatrix}
    1 & 0\\
    0 & 1
    \end{pmatrix}}
\def \rodue{ \frac 1{\sqrt 2}
    \begin{pmatrix}
    0 & a\\
    b & 0
   \end{pmatrix}}

\def \Rank{\text{Rank}}

\begin{document}

\title{Entanglement entropy of the long-range Dyson hierarchical model }
\author{Silvia Pappalardi$^{1,2,3}$, Pasquale Calabrese$^{1,2,4}$, Giorgio Parisi$^{5}$}
\address{$^1$ SISSA, Via Bonomea 265, I-34136 Trieste, Italy}
\address{$^2$ Abdus Salam ICTP, Strada Costiera 11, I-34151 Trieste, Italy}
\address{$^3$ Department of Physics, Boston University, 590 Commonwealth Avenue, Boston, Massachusetts 02215, USA}
\address{$^4$ INFN, Sezione di Trieste,  I-34151 Trieste, Italy}
\address{$^5$ Dipartimento di Fisica,
  Universit\`a di Roma la Sapienza, INFN, Sezione di Roma 1, and CNR-Nanotec,
  I-00185 Rome, Italy}


\begin{abstract}
We study the ground state entanglement entropy of the quantum Dyson hierarchical spin chain in which the interaction decays 
algebraically with the distance as $r^{-1-\sigma}$.
We exploit the real-space renormalisation group solution which gives the ground-state wave function in the form of a tree tensor network and 
provides a manageable recursive expression for the reduced density matrix of the renormalised ground state.
Surprisingly, we find that at criticality the entanglement entropy obeys an area law,  as opposite to the logarithmic scaling of short-range critical systems and of other
non-hierarchical long-range models. 
We provide also some analytical results in the limit of large and small $\sigma$ that 
are tested  against the numerical solution of the recursive equations. 
\end{abstract}

\maketitle

\section{Introduction}
\label{sec:introduction}

Long-range interacting systems, because of their unconventional static and dynamical properties, have been extensively studied in classical statistical mechanics since many decades \cite{campa2014physics}. 
In the quantum domain, atomic, molecular and optical systems (AMO), as well as synthetic ones, are often described by long-range interactions. 
Yet, they have been a focus of great attention only relatively recently, as a consequence of their experimental implementation in different physical setups of many-body AMO systems,
such as Rydberg atoms \cite{saffman2010quantum, schauss2012observation}, dipolar molecules \cite{yan2013observation, lu2012quantum} or trapped ions \cite{Albiez2005m, blatt2012quantum, jurcevic2017direct, neyenhuis2017observation}. 
In particular trapped ions, which interact via ferromagnetic Ising long-range interactions \cite{haffner2008quantum, pagano2018cryogenic},
represent one of the most promising platforms for quantum computation. 
These experimental advances motivated an increasing interest towards the understanding of quantum systems with long-range interactions 
both in thermodynamic equilibrium \cite{v-96,lab-05,dutta2001phase,maghrebi2016causality,defenu2017criticality,gpsst-15, koffel2012entanglement,frerot2017entanglement,fs-16,jki-14, lvp-16,zhe2017entanglement, ares2019entanglement,gr-12,gr-13,aefz-19,vodola2014kitaev,gabbrielli2018multipartite, roy2018effect,vvbv-18,hgca-17} 
and in several non-equilibrium situations
\cite{ht-13,eisert2013breakdown,zmk-15,gr-14,rs-15,tgc-14, fnr-17, smr-17,hv-14,cevolani2016spreading,cdcts-18, tran2018locality,pappalardi2018scrambling, lerose2018quasi,btsr-17,zunkovivc2018dynamical, jaschke2017critical, Daley,DaleyEssler,lerose2018logarithmic,gc-18,hz-17,dekm-18}.
Anyhow, there are still only few results concerning the equilibrium properties of quantum spin chains with ferromagnetic power-law decaying interactions, 
see e.g. Refs. \cite{dutta2001phase, maghrebi2016causality, defenu2017criticality} for some mean-field and perturbative solutions and 
Refs. \cite{ jaschke2017critical,werner2005phase,werner2005quantum,sss-12} for some numerical simulations. 
Even for the long-range ferromagnetic Ising model, the precise knowledge of the phase-diagram and universal features at the 
quantum phase transition are still lacking.

At the same time, it is by now well established that a large amount of information about many-body systems can be inferred by their entanglement 
properties \cite{amico2008entanglement,calabrese2009entanglement,laflorencie}. 
The most useful quantity is surely the celebrated \emph{entanglement entropy}, defined as follows. 
Let us consider a quantum system in a pure state with density matrix $\ro = \ket{\psi}\bra{\psi}$ and a bipartition of the 
Hilbert space $\mathcal H = \mathcal H_A \otimes \mathcal H_B$, 
the entanglement between subsystems $A$ and $B$ may be measured by the Von Neumann entropy of the reduced density matrix 
$\ro_A = \Tr_{B}{\hat \rho}$, i.e. 
\begin{equation}
\label{eq:EE}
S_A =  - \Tr\,[ \ro_A \ln \ro_A] \ .
\end{equation}
The entanglement entropy turned out to be a very useful quantity, especially when $A$ is a block of $\ell$ contiguous lattice sites in a one dimensional chain.
In fact, the scaling entanglement entropy is tightly connected to the presence of quantum phase transitions for short-range systems. 
Indeed, while for gapped systems with short-range interaction, $S_{\ell}$ obeys an area law \cite{sredniki,eisert-2010,hastings2007area} 
(i.e. it scales  with surface of the boundary between 
the two partitions), it  grows like   $\ln \ell$ at critical point, with a proportionality constant given by the central charge of the underlying conformal field theory \cite{hlw-94,cc-04, vidal2003entanglement,cc-09}. 

However, for long-range quantum interactions conformal invariance is broken  as a consequence of the lack of Lorentz symmetry (rotational one in Euclidean spacetime) 
as explicitly manifested by a dynamical critical exponent $z \neq 1$ \cite{dtr-16}. 
Hence unusual ground state entanglement properties are expected \cite{koffel2012entanglement, gpsst-15,frerot2017entanglement,  zhe2017entanglement, ares2019entanglement,gr-12,gr-13,aefz-19}, as e.g. shown numerically for antiferromagnetic long-range one-dimensional systems, which 
violate area-law scaling  also in the gapped phase \cite{koffel2012entanglement,vodola2014kitaev,gabbrielli2018multipartite, roy2018effect}. 
On the other hand, the ferromagnetic mean-field Ising model has a critical entanglement entropy that scales with the logarithmic of the number of spins in $A$ 
\cite{latorre2005entanglement,vidal2007entanglement}.  
Beyond mean field, the scaling of entanglement entropy for ferromagnetic long-range interaction  is still an open issue, also in one spatial dimension. 

In this work, we address such a problem, by considering the quantum version of the ferromagnetic  Dyson hierarchical model \cite{monthus2015dyson, monthus2015dysonGlass}. 
Its classical version was introduced by Dyson in 1969 \cite{dyson1969existence, dyson1971ising} and it provided, via an exact renormalisation group (RG) approach, analytical insights about the critical behaviour  of long-range interacting one-dimensional spin chains. 
Also the quantum version has been recently solved by Monthus \cite{monthus2015dyson}
using real-space renormalisation and the 
critical  properties of the ground state have been obtained by a recursive projection onto the low lying energy states. 

Within this frame, we study the entanglement entropy of the renormalised ground state of the Dyson hierarchical model at criticality. 
The result depends on the choice of the subsystem because the model is not translationally invariant. 
Hence, we consider three types of partitions and, for each of them, we determine recursively the reduced density matrix. 
We finally analyse  the scaling of the entanglement entropy at the critical point and in some limits we derive analytical expansions. 
Our main and surprising result is that, although the correlation functions decay algebraically, 
the entanglement entropy obeys an area law. 
This is due to the hierarchical structure of the renormalised  ground state that makes the rank of the reduced density matrix finite
(the ground state is a tree tensor network \cite{tagliacozzo2009simulation,mvlm-10,alba2011entanglement,t-prep} with finite bond dimension). 
In spite of this saturation at the critical point, the renormalised  ground-state reduced density matrix reproduces the power-law decay of the two-point
correlation function with the exact critical exponent. 
 
The paper is organised as follows. 
In Sec. \ref{Sec:model}, we  describe the model and the main steps of the real-space RG procedure. 
Sec. \ref{sec:results} contains all our results, including  a description of the three considered partitions and a first discussion of our findings. 
This section is divided as:
in Subsecs. \ref{sec:power2}-\ref{subsec:two_cut} we derive recursively the reduced density matrix and the entanglement entropy for the three considered partitions;
 we then show by an elementary approach that the rank of the reduced density matrix is finite (Subsec. \ref{sec:general_beha})
and that  the correct power-law decay of the correlation functions follows from them (Subsec. \ref{sec:corr}).
In the closing section we draw our conclusions and discuss some open questions.
In the appendix we report some technical details of the calculations. 

\section{The model}
\label{Sec:model}

The Dyson hierarchical model was introduced by Dyson in 1969 \cite{dyson1969existence, dyson1971ising} with the aim of studying phase transitions in one dimensional ferromagnetic Ising models with algebraically decaying interactions with exponent $1+\sigma$. 
A hierarchical Hamiltonian is defined with a tree structure on a spin chain of length $L=2^n$, where blocks of spins are subject to an interaction that decays as a power-law of their distance. 
In the classical case, the hierarchical model reproduces well the critical properties of the long-range Ising Hamiltonian \cite{parisi1988stati}. 
In fact, for $\sigma<1/2$ the critical exponents of  the former coincide with the mean-field of the latter. 
Furthermore, for $\sigma\geq 1/2$, the exponents of the two models coincide at first order in the $\epsilon=\sigma-1/2$ expansion  \cite{Collet1978renorma}.

\begin{figure}[t]
\centering
\fontsize{13}{10}\selectfont
\includegraphics[width=0.98\textwidth]{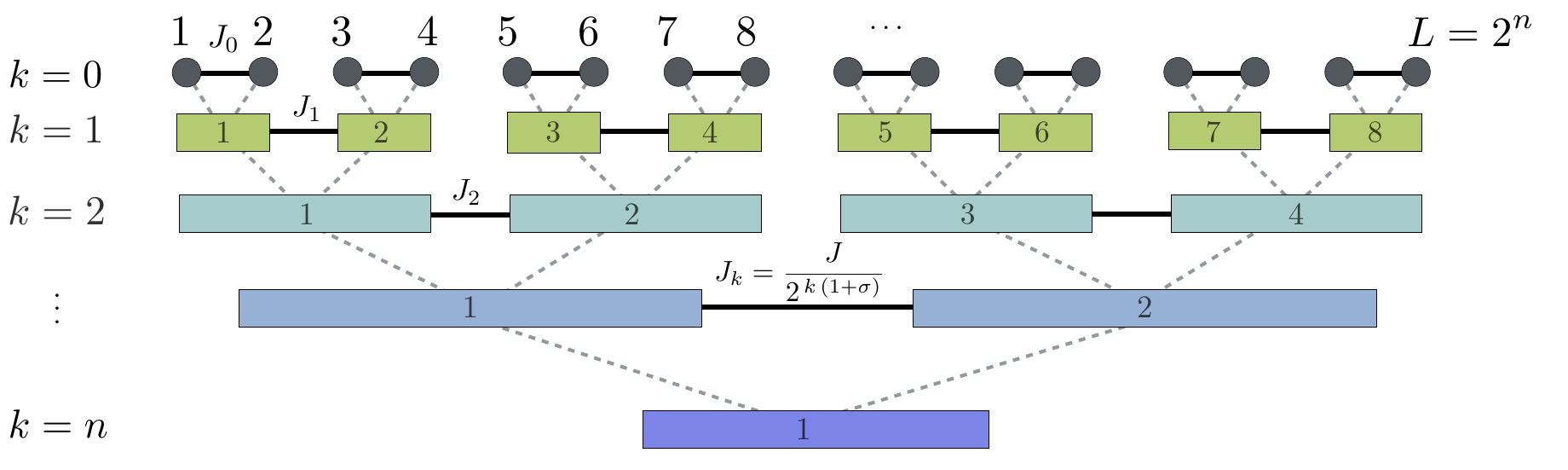}
\caption{Tree structure of the Dyson hierarchical model. The Hamiltonian is a sum over different two-body interactions $J_k$ between block variables at level $k$
(full horizontal lines). 
The level $k=0$ corresponds to the physical spins (top circles). At each layer, the block variables, represented by coloured boxes, 
interact with the coupling $J_k=J/2^{k(1+\sigma)}$. The dashed lines show which spins at level $k$ form the blocks at level $k+1$.
The real-space RG procedure consists in replacing each box with an effective spin. 
At the last step, there is a single spin left and the Hamiltonian is an effective transverse magnetic field $h_n$. 
}
\label{fig:tree}
\end{figure}

The quantum version of this model has been introduced and studied with real-space RG by Monthus in \cite{monthus2015dyson, monthus2015dysonGlass}. 
Let us describe the Hamiltonian which is pictorially represented in Fig. \ref{fig:tree}. 
At step $k=0$ each pair of spins at sites $2j-1$ and $2j$ interacts with a coupling $J_0=J$  as $J \hat \sigma_{2j-1i}^{x} \hat \sigma_{2j}^{x}$,
where hereafter $\hat \sigma_i^{\alpha}$ denotes the Pauli matrices.
The absence of interactions between the sites $2j$ and $2j+1$ is the main difference from standard Ising models. 
At the following step $k=1$, two blocks of two spins interact (always along $x$) with coupling $J_1= J/2^{1+\sigma}$, then at $k=2$ blocks of four spins interact with $J_2= J/2^{2(1+\sigma)}$ and so on. 
Hence, the total hamiltonian is 
\begin{align}
\label{eq:blok_interaction}
 \hat H &=- \sum_{k=0}^{n-1} J_k\, \mu^2
    \sum_{i=1}^{L/2^{k+1} } \Big( \sum_{j=0}^{2k} \hat \sigma^x_{2^k(2i-1)-j} \Big) \Big( \sum_{j=0}^{2k} \hat \sigma^x_{2^{k+1}i-j} \Big)
 -  h\, \sum_{i=1}^{L} \hat{\sigma}_i^z, \quad 
\end{align}
%
%
where $J_k= {J}{2^{-k (1+\sigma)}}$ and 
we introduced the transverse field $h>0$ and the magnetic moment $\mu $ (the latter will be set to 1, and it has been introduced for convenience in the renormalisation procedure). 
The real-space RG of Ref. \cite{monthus2015dyson}  is constructed by projecting (at each step $k$) the Hamiltonian $\hat H^{(k)}$ onto the two lower energy states. 
 Hence at each RG step, we have half of the spins. 
The renormalised  transverse field $h_{k}$ and magnetic moment $\mu_{k}$ are obtained within this procedure.

\subsection{Real-space RG for the Dyson hierarchical model}
\label{Sec:RG_procedure}
We report here the main ideas and the conceptual steps of the real-space RG approach. 
For all the details we refer to \cite{monthus2015dyson}. 
We rewrite the Hamiltonian (\ref{eq:blok_interaction}) 
as a sum over the generations $k=0,1,..,n-1$  
\begin{equation}
\hat H  = \sum_{k=0}^{n-1} \hat H^{(k)} \ ,
\label{recDyson}
\end{equation}
where $\hat H^{(k)}$ for $k>1$ is read off Eq. (\ref{eq:blok_interaction}), and in particular  $k=0$ is 
\begin{equation}
 \hat H^{(k=0)}  
  =
 - h\, \sum_{i=1}^{L} \hat {\sigma}_i^z
 - \sum_{i=1}^{L/2}  J_{0} [\mu \, \hat{ \sigma}_{2i-1}^x] [ \mu \,\hat{\sigma}_{2i}^x] 
  = \sum_i^{L/2}\, \hat H^{(0)}_{2i}
 \ .
\label{h0dyson} 
\end{equation}
The Hamiltonians $\hat H^{(0)}_{2i}$ are unrelated and each pair of spins may be diagonalised  independently. 
It is convenient to work in the basis of eigenstates of  $\hat\sigma^z_i$, i.e. $\hat \sigma^{z}_{i}\ket{\pm}_{2i}= \pm \ket{\pm}_{2i}$. 
Then, the four eigenstates and eigenvalues of $\hat H^{(0)}_{2i}$ are
\begin{subequations}
\label{eq:lowest}
\begin{equation}
\,\, \ket{\lambda^{+}_{\text{s/a}}}_{2i} = a_{1} \ket{+}_{2i-1}\ket{+}_{2i}
     \pm b_{1}\, \ket{-}_{{2i-1}}\ket{-}_{2i} \ ,\quad 
     \lambda^+_{\text{s/a}} = \mp \sqrt{(J_0\mu^2)^2+4h^2 } \ ,
\end{equation}
\begin{equation}
\ket{\lambda^{-}_{\text{s/a}}}_{2i} = \frac 1{\sqrt 2}   \ket{-}_{2i-1}\ket{+}_{2i}
        \pm \frac 1{\sqrt 2}\,  \ket{+}_{2i-1}\ket{-}_{2i} \ , \,          \quad 
       \lambda^-_{\text{s/a}} = \mp J_0\, \mu^2 \ , \quad\quad \quad
\end{equation}
\end{subequations}
where ``s/a'' stands for the symmetric and anti-symmetric combination respectively and the coefficients are
${{a_1 = \sqrt{1/2+{h}/ \sqrt{ (J_0 \mu)^2+4 h^2 }}}}$, $b_1=\sqrt{1-a_1^2}$. 
In the RG approach, we keep only the two lowest energy states $\ket{\lambda^{\pm}_{\text{s}}}_{2i}$.
Thus the RG transformation is the  projection onto the symmetric subspace. 
The \emph{renormalised  spins} at level $k=1$ are then identified by the basis 
\begin{equation}\label{eq:identi}
\ket{\pm}^{[1]}_{{2i}} \equiv \ket{\lambda^{\pm}_{\text{s}}}_{2i} \ .
\end{equation}
They are related to the physical spins by Eq. (\ref{eq:lowest}) and are pictorially represented in Fig. \ref{fig:tree} by the green boxes. 
Hereafter, the superscript $[k]$ always refers to a renormalised spin at level $k$. 

Eq. (\ref{eq:identi}) allows us to define the projection operator 
\begin{equation}\label{eq:projet}
\hat P_1 \equiv \prod_i \hat P_{1,\,2i} = \prod_i \, \ket{+} \bra{+}^{[1]}_{2i} + \ket{-}\bra{-}^{[1]}_{2i} \ ,
\end{equation}
which implements the RG transformation and selects the two lowest energy states of $\hat H^{(0)}$. 
When applied to $\hat H^{(0)}+ \hat H^{(1)}$ in Eqs. (\ref{recDyson}-\ref{h0dyson}), the projection gives the renormalised  Hamiltonians
\begin{equation}\label{eq:reno}
\hat H_R^{(0)}+ \hat H_R^{(1)} = 
 - \frac{L}2 e_1\, \hat P_1 - h_1 \sum_{i}\, \hat \sigma^{z\,[1]}_{2i} 
- \sum_{i=1}^{L/4}  J^{(1)} [\mu_1 \, \hat{ \sigma}_{4i-2}^{x\,[1]}] [ \mu_1 \,\hat{\sigma}_{4i}^{x\,[1]}]  \ ,
\end{equation}
where $h_1$, $\mu_1$ are the renormalised  magnetic field and moment while $e_1$ is the contribution of the previous generation to the ground state energy. 
It is then clear that the RG flow should be written in terms of the control parameter
\begin{equation}\label{Eq:flow}
K_k \equiv \frac{J_k\, \,\mu^2_k}{h_k} , \qquad  \text{with} \quad K_0 = \frac{J_0\, \mu^2}{h}= \frac J h \ .
\end{equation}
Hence, at level $k=1$ the renormalised parameter are
\begin{subequations}\label{eq_recu1}
\begin{align}
h_1 & = \frac{\lambda^-_{\text s}-\lambda^+_{\text s}} 2 \ = 
    -\frac {2 h}{ K_0\, +  \sqrt{K_0^2+4}}\ ,\\
e_1 & = \frac{\lambda^-_{\text s}+\lambda^+_{\text s}}2  = -h\, \frac{K_0+\sqrt{K_0^2+4}}2 \ ,  \\
\mu_1 & = \mu \sqrt{2}\sqrt{1+\frac{K_0}{\sqrt{K_0^2+4}} }  \ .
\end{align}
\end{subequations}
In Eq. (\ref{eq:reno}) the operators $\sigma^{z/x\,[1]}_{2i}$ are Pauli matrices acting on the renormalised spins at level $k=1$ as $\sigma^{z\,[1]}_{2i}\ket{\pm}^{[1]}_{2i}= \pm \ket{\pm}^{[1]}_{2i}$ and $\sigma^{x\,[1]}_{2i}\ket{\pm}^{[1]}_{2i}= \ket{\mp}^{[1]}_{2i}$. 
The renormalised Hamiltonian (\ref{eq:reno}), besides the projector $\hat P_1$, has exactly the same structure of $\hat H^{(0)}$ in Eq. (\ref{h0dyson}), 
being a sum of independent two-spin Hamiltonians. 
Eqs. (\ref{eq_recu1}) are the renormalisation rules for the magnetic field, coupling, and energy. 

The diagonalisation and projection transformations (\ref{eq:lowest}-\ref{eq:reno}) are then applied again $n-1$ times on all the terms of the Hamiltonian (\ref{recDyson}).
The $k$-th renormalised spin is the  block of $2^k$ spins at level $k$. 
Its basis $\ket{\pm}_i^{[k]}$ is defined recursively as
\begin{subequations}
\label{eq:recu+}
\begin{align}
\ket{+}^{[k]}_i & = a_{k-1} \ket{+ \,+\,}^{[k-1]}_i +b_{k-1}\, \ket{- \,-\,}^{[k-1]}_i \ , \\
\ket{-}^{[k]}_i & = \frac 1{\sqrt 2} \, \ket{- \,+\,}^{[k-1]}_i  + \frac 1{\sqrt 2}\,  \ket{+ \,-\,}^{[k-1]}_i \ .
\end{align}
\end{subequations}
Here $\ket{a \,b\,}^{[k-1]}_i$ stands for the tensor product of two adjacent $(k-1)$-th renormalised spins  as 
${\ket{a \,b\,}^{[k-1]}_i= \ket a_{2i-1}^{[k-1]}\otimes \ket b_{2i}^{[k-1]}}$. 
The parameters $a_k$, $b_k$ are obtained from the diagonalisation and read
\begin{equation} \label{eq:state_coeff}
a_k = \frac 1{\sqrt 2} \left(1 + \frac 2{\sqrt{K_k^2 + 4} }
    \right)^{1/2} 
    \ , \quad \quad \quad
b_k = \frac 1{\sqrt 2} \left(1 - \frac 2{\sqrt{K_k^2 + 4} } 
    \right)^{1/2}  \ ,
\end{equation}
while the contribution of each projection to the ground state energy is
\begin{equation}
 e_{k} =- h_k \frac{K_k+ \sqrt{ K^2_k + 4 }  }{2} \ .
\label{rgegspur}
\end{equation}
At every step, the renormalisation rules   
\begin{eqnarray}\label{eq:reno_rules}
h_{k+1}  =   \frac{ 2 h_{k} }
{K_k + \sqrt{ K_k^2+4 } }
\label{hrpur} \ , \quad \quad 
\mu_{k+1} 
 =\mu_k \sqrt{2}   \sqrt{ 1+ \frac{K_k} {\sqrt{ K_k^2+4 }}}  \ ,
\end{eqnarray}
give explicitly the recursion relation for the control parameter (\ref{Eq:flow}) as 
\begin{equation} \label{eq:flow_rg}
K_{k} = \frac{ K_{k-1}}{2^{1+\sigma}}\, 
    \frac {\left (\,K_{k-1} + \sqrt{K_{k-1}^2 + 4} \, \right )^2}{\sqrt{K_{k-1}^2 + 4} }
    \equiv R[K_{k-1}] \ .
\end{equation}
%
%

\subsection{The RG ground state}
After renormalisation, the hierarchical Hamiltonian (\ref{recDyson}) becomes
\begin{equation}
\label{eq:reno_ham}
\hat H_R = - \sum_{k=1}^{n} \frac{L}{2^k}e_{k} \, \hat P_k - h_n\, \hat \sigma^{z\, [n]}  ,
\end{equation}
where $e_k$ is the energy contribution (\ref{rgegspur}) of each renormalisation step, 
$\hat P_k$ is the projection operator onto the lowest energy eigenstates at level $k-1$ (i.e. the generalisation  of Eq. (\ref{eq:projet})), 
$h_n$ is the renormalised transverse  field (\ref{eq:reno_rules}) and $\hat \sigma^{z\,[n]} $ is the Pauli matrix at level $n$. 
Hereafter, we will not use the subscript for the Pauli matrices and eigenvectors at level $n$ because there is a single spin at that level. 
Since $h_n>0$, the ground state is simply 
\begin{equation}
\label{eq:gs}
\ket{\psi_{\text{GS}}}  = \ket{+}^{[n]}  ,
\end{equation}
and it can be written in the basis of the spins at the previous level using Eq. (\ref{eq:recu+}) as
\begin{equation}
\label{eq:gs_recu}
\ket{\psi_{\text{GS}}} = \ket{+}^{[n]} = a_{n-1} \ket{+ \,+\,}_1^{[n-1]} +b_{n-1}\, \ket{- \,-\,}_1^{[n-1]} \ ,
\end{equation}
and the recursively up to writing it in the physical spins at level $0$.

Note that not only the Hamiltonian has a binary tree form, but also the RG ground state has the structure of a \emph{tree tensor network} 
where the isometries have all rank equal to two, see \cite{tagliacozzo2009simulation,mvlm-10, alba2011entanglement,t-prep}. 
In fact, the transformations of Eq. (\ref{eq:recu+}) automatically define the isometric tensors in the sense of Ref. \cite{tagliacozzo2009simulation}.

\subsection{The critical point}
The critical point and all the universal exponents are determined from the study of the flow induced by the RG equation (\ref{eq:flow_rg}). 
The critical point $K_c=R[K_c]$ depends on $\sigma$. 
For $\sigma \ll 1$,  $K_c\to 0$ (aka $h_c$ diverges) and all the spins align along the $z$ direction. 
Conversely, for $\sigma \gg 1$, $K_c\to \infty $ (aka $h_c\to 0$).
Hence, as in the usual quantum Ising model \cite{sach-book}, the critical point separates a ferromagnetic phase, for $h<h_c$ or $K>K_c$, 
from a paramagnetic one, for $h>h_c$ or $K<K_c$.
At the critical point, the ground-state wave function simplifies considerably. 
Since $K_k=K_c$ for all $k$, the coefficients of Eq. (\ref{eq:state_coeff}) do not depend on $k$, i.e.  $a_k=a\,,b_k=b$ for all $k$
 and the recurrence relation (\ref{eq:gs_recu}) further simplifies, as we will exploit to calculate the entanglement entropy.

By linearising  the flow equation (\ref{eq:flow_rg}) close to the fixed point, the critical exponents for the correlation length ($\nu_\sigma$), correlation function ($x_\sigma$), 
and the dynamical one ($z_{\sigma}$) have been determined \cite{monthus2015dyson}. 
We will be interested in the longitudinal correlation function between two spins at distance $r$, that 
at the critical point, behaves as
\begin{equation}\label{eq:corre_fu}
\langle \hat \sigma^x_i\, \hat \sigma^x_{i+r} \rangle \sim \frac 1{r^{2x_{\sigma}}}\ ,
\end{equation}
where the exponent $x_{\sigma}$  is \cite{monthus2015dyson}
\begin{equation}
x_{\sigma} = \frac{1-\sigma}{4}+ \frac{\ln(K_c^2+4)}{8 \ln 2} \ .
\label{xpur}
\end{equation}
We mention that in the limit $\sigma \ll 1$, all the critical exponents in \cite{monthus2015dyson} reproduce the mean-field results of \cite{dutta2001phase, maghrebi2016causality}. 
For larger $\sigma$, they are in good qualitative agreement with the values obtained numerically in a dissipative short-range model in the same universality class of the 
long-range Ising one\cite{werner2005phase, werner2005quantum,sss-12}.


\section{The block entanglement entropy of the hierarchical model}
\label{sec:results}

\begin{figure}[t]
\centering
\fontsize{13}{10}\selectfont
\includegraphics[width=0.98\textwidth]{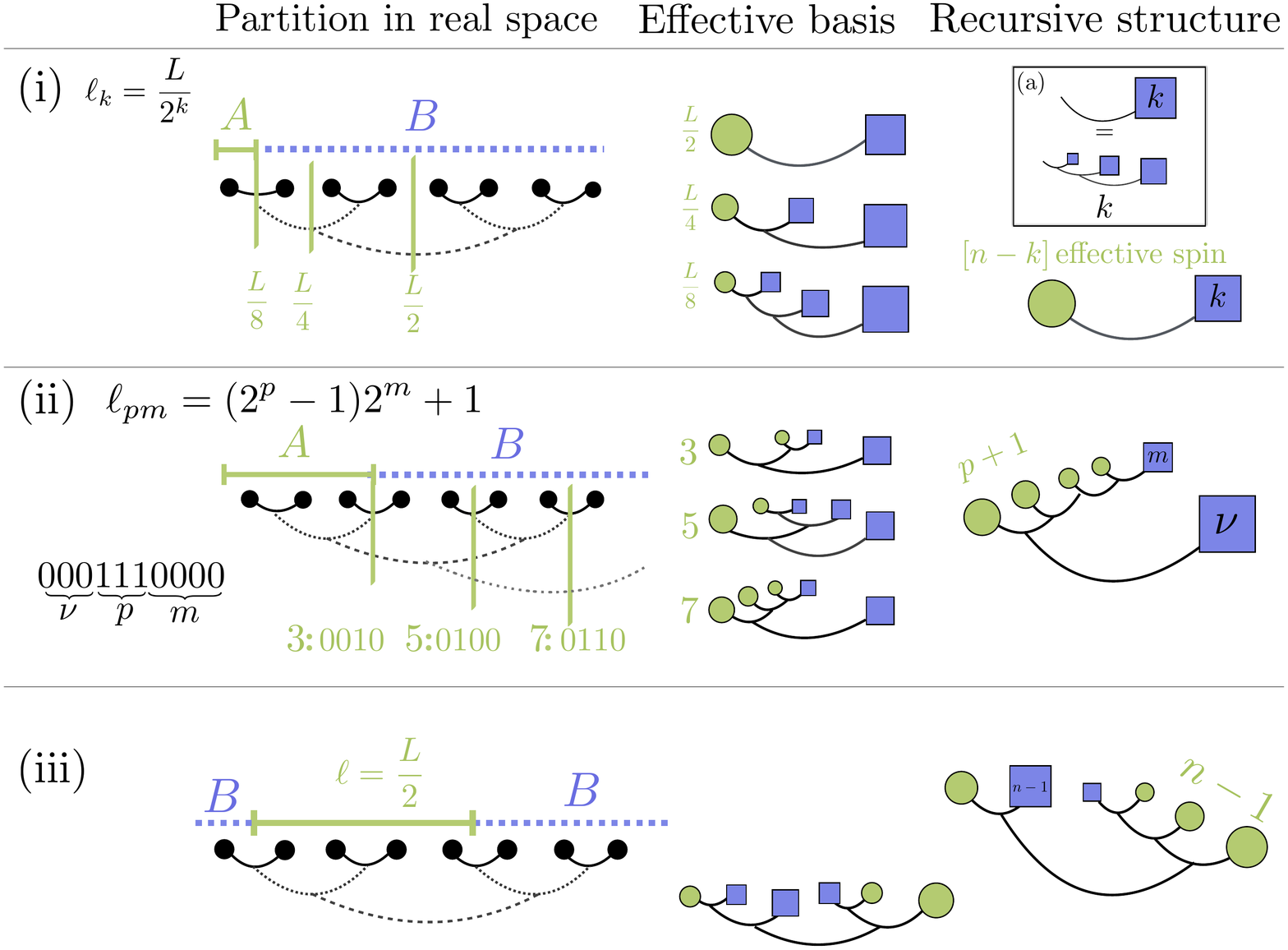}
\caption{The partitions (\ref{sec:0}-\ref{sec:1}) of the ground state of the hierarchical model that we consider here. 
Left column: the partitions in real space. 
Central column: reduced  density matrix in renormalised variables. The green circles indicate the block spins, whose basis is used to write the recursive density matrix. 
Each of them contributes with a factor two to the size of $\ro_{A}$. 
Right column: recursive structure of the reduced density matrix. 
In the inset (a) we introduce the diagram for the recursive trace performed on a block-spin $k$ times.
(Top row) Partition (\ref{sec:0}) with an interval of size $\ell_k=2^{n-k}$. 
The reduced density matrix is a $2\times 2$-matrix in the basis of the renormalised spins at $(n-k)$-th level.  
(Middle) Partition (\ref{Sec:2}) with an interval of length $\ell_{pm}=(2^p-1)2^m+1$. 
The reduced density matrix has effective size $2^{p+1}\times 2^{p+1}$.  
(Bottom) Partition (\ref{sec:1}) with an interval of length $\ell=L/2$ shifted by one site from the chain end. 
This partition always splits $n$ effective spins and the reduced density matrix has size $L\times L$.}
\label{fig:tree_scaling}
\end{figure}

In this section we present the main results of this paper about the entanglement entropy of the RG ground state of the Dyson hierarchical model at the critical point. 
We mention that a very large literature exists  about the characterisation of the entanglement in many-body systems by real space RG, but its main focus is disordered 
systems and strong disorder RG \cite{rm-04,rm-09,s-06,fcm-11,lir-07,il-08,ki-12,dmh-17,rrs-14,rac-16,joz-18,randbow}, see \cite{im-18} as a recent review. 
Although the employed techniques are different, there are many qualitative similarities that also helped our understanding of the hierarchical model.

We consider bipartitions of the chain into a block of spins $A$ of length $\ell$ and the remainder $B$.
The first step will be to reconstruct the reduced density matrix $\ro_A$ of the ground state (\ref{eq:gs}) as achieved by means of the recursive projection technique 
in the previous section. We then calculate the eigenvalues of $\ro_A$ and consequently the entanglement entropy (\ref{eq:EE}). 
Since local transformations within the subsystems do not alter the entanglement between $A$ and $B$, 
we will write $\ro_A$ without tracing back all those renormalised spins that are not split by the bipartition: in tree tensor network 
language \cite{tagliacozzo2009simulation,t-prep}, 
this means that all the isometries contained within $A$ do not contribute to the entanglement  and cancel in the construction of the reduced density matrix. 
The size of the reduced density matrix is then determined by the number of the renormalised spins (\ref{eq:recu+}) that are split by the bipartition. 
As a simple example, let $A$ be the block composed of the first $\ell= L/2$ sites. 
This partition spits only the $n$-th effective spin at level $n$, as in the top panel of Fig. \ref{fig:tree_scaling} for $k=1$. 
Hence, the reduced density matrix $\ro_A$ is a $2\times 2$ matrix in the basis $\ket{\pm}^{[n-1]}_1$ and its eigenvalues will never change when 
rewriting in terms of the spins at lower levels up to the physical spins. 

The same reasoning applies to all other kinds of intervals. However,
since the model is not translationally invariant, the reduced density matrix depends not only on the length of the subsystems $A$, but also on its position in the chain.
We consider the three following bipartitions as shown in Fig. \ref{fig:tree_scaling}: 
\begin{enumerate}
\item the interval of length $\ell_k =L/2^k= {2^{n-k}}$ (with $k=1,\dots n$) starting from left end of the chain as in the top panel of Fig. \ref{fig:tree_scaling}; \label{sec:0}
\item the interval of length $\ell_{pm}=(2^p-1) 2^m+1$ (with $m,p\geq 1$) again  starting from left end of the chain as in the central panel of Fig. \ref{fig:tree_scaling}; \label{Sec:2}
\item the shifted interval of length $\ell=L/2$ starting from the second site of the chain as in the bottom panel of Fig. \ref{fig:tree_scaling}. \label{sec:1}
\end{enumerate}
For these three cases, we will construct the reduced density matrix $\ro_{\ell}$ recursively.

The partition (\ref{sec:0}), as previously discussed for $\ell=L/2$, splits only one renormalised spin at level $n-k$.
Hence, the reduced density matrix is conveniently written in the basis at this level when it is again a $2\times 2$ matrix, whose elements will be obtained recursively. 

Partition (\ref{Sec:2}) instead cuts $p+1$ effective spins, hence it is a matrix of size $2^{p+1}\times2^{p+1}$ in the basis given by the tensor product of the effective spins at corresponding levels (see below for details). 
Consequently the largest possible rank is obtained for $\ell=L/2-1$, i.e. $m=1$ and $p=n-2$.

The partition (\ref{sec:1}) is chosen in such a way to have maximum possible rank (as standard in tree tensor 
networks \cite{tagliacozzo2009simulation,mvlm-10, alba2011entanglement,t-prep}). 
Indeed, this partition cuts by construction $n$ effective spins and so $\ro_A$ is, in principle, a $L\times L$ density matrix.

\begin{figure}[H]
\centering
\fontsize{12}{10}\selectfont
\includegraphics[width=0.676\textwidth]{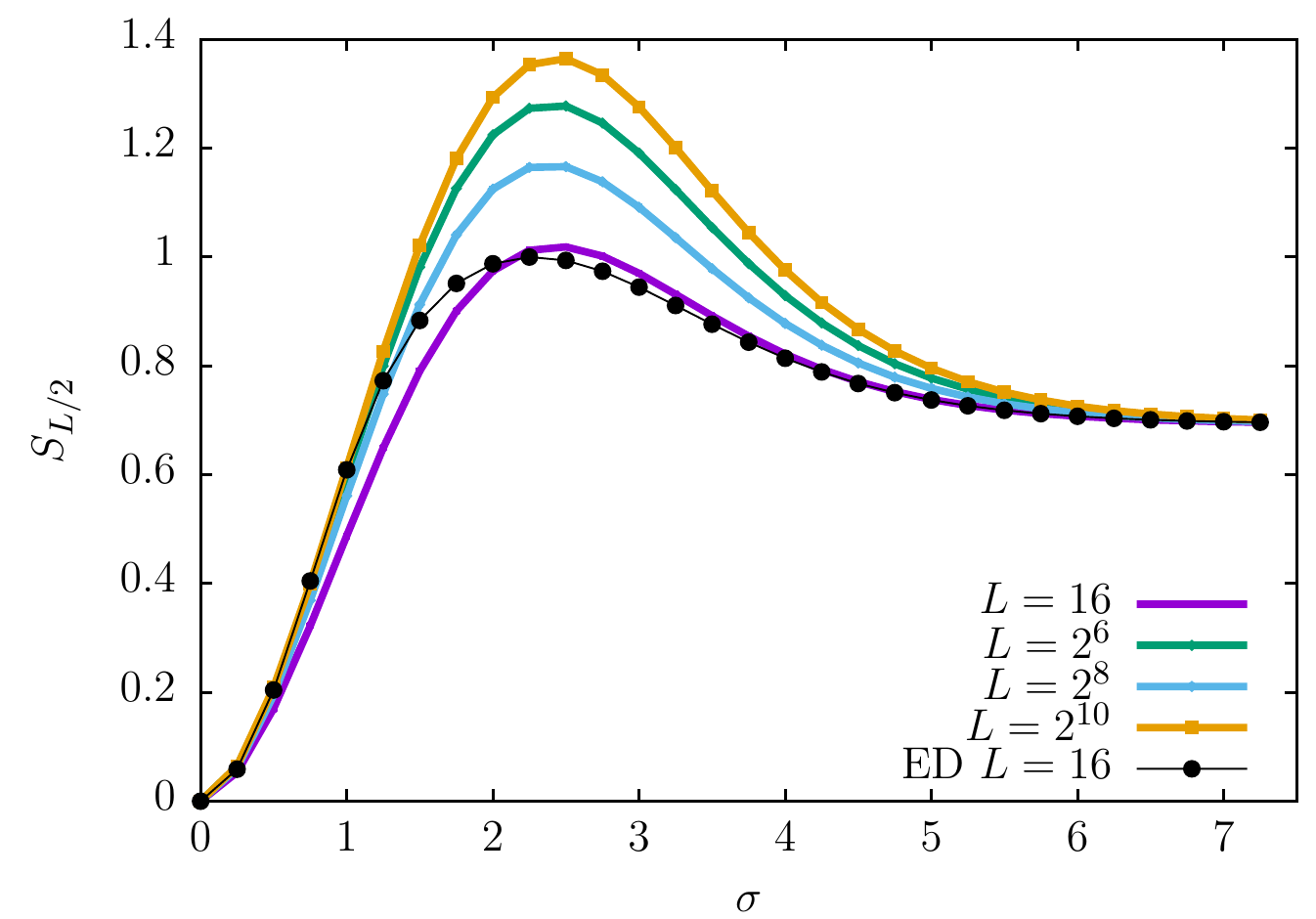}
\caption{Entanglement entropy at the critical point for the partition (\ref{sec:1}) as a function of $\sigma$.  
$S_A$ is non-monotonic. 
We compare numerical exact diagonalisation results at  $L=16$ with the RG prediction. 
For intermediate $\sigma$, finite size effects are relevant.}
\label{fig:one_cutEE_sigma}
\end{figure}

Numerical and analytical results for the entanglement entropy at criticality will be explicitly obtained in the following subsections for asymptotically large systems and subsystems.
The most relevant and surprising result is that the reduced density matrix has alway a finite rank (at most equal to 16), even for the case (\ref{sec:1}) 
when the rank naively might have been maximal, i.e. $L$. 
The natural question at this point is how well the fixed point ground-state entanglement correctly describes the one of the real model, without integrating out the 
high-energy physics in the renormalisation procedure.
To this aim, we  compare the renormalised entanglement with the exact one.
In Fig. \ref{fig:one_cutEE_sigma} we report the comparison with the data from numerical exact diagonalisation for $L=16$ and partition (\ref{sec:1})
(because of the tree structure, the next possible size would be $L=32$ that is hard to exactly diagonalise; 
alternatively one can use tree tensor networks to get very accurate estimates for larger chain, but this is beyond the goal of this paper).
We show the comparison for many values of $\sigma$.
The two curves are qualitatively very similar, although  $S_A$ is non-monotonic in $\sigma$. 
While for large and small $\sigma$ they compare very well, there are quantitative differences for $\sigma\sim1$.
Most probably this is just a finite size effect and indeed it is remarkable that for such small value of $L$, the two curves match so well. 
In the figure we report also the results from renormalisation group for larger $L$, showing that the saturation value is much larger for intermediate 
$\sigma$ as we shall discuss.  

Since in the following we will only consider critical entanglement entropies, 
to lighten the notations we set $K_c=K$ to indicate the control parameter at the critical point and we set $J=1$.

\subsection{Entanglement entropy for the partition (\ref{sec:0}).}
\label{sec:power2}
We first consider the reduced density matrix for an interval $A$ of length $\ell_k=2^{n-k}$ starting from the  chain boundary.
The partial trace over $B$ (the complement of $A$, i.e. the last $L-\ell_k$ spins) may be obtained recursively by tracing away half of the spins at each step $k$ times, 
as shown in the top panel of Fig. \ref{fig:tree_scaling}. 
For later convenience, it is useful to define the auxiliary matrices $\hat \tau_i^{[k]}$, represented pictorially by the diagram $(a)$ in the inset of Fig. \ref{fig:tree_scaling},  as
\begin{subequations}
\label{eq:alpha_def}
\begin{equation}
\hat \tau_0^{[k]}  \equiv \tr_{B} \ket {+} \bra {+}^{[n]}  \ , \quad 
\hat \tau_1^{[k]} \equiv \tr_{B} \ket {-} \bra {-}^{[n]} \ , 
\end{equation}
\begin{equation}
\hat \tau_2^{[k]} \equiv \tr_{B} \ket {+} \bra {-}^{[n]} \ , \quad 
\hat \tau_3^{[k]} \equiv \tr_{B} \ket {-} \bra {+}^{[n]}  \ ,
\end{equation}
\end{subequations}
where the states $\ket{\pm}^{[n]}$ are defined in Eq. (\ref{eq:recu+}). Notice that the $k$ dependence is encoded in $B$. 
These matrix can be written explicitly in the basis of the first spin at level $n-k$ (i.e. in the basis $\ket{\pm}^{[n-k]}_1$) by recursion (see \ref{app:prood_powe_two}) as
\begin{equation}
\label{eq:alpha_basis}
\hat \tau_0^{[k]}  = 
    \begin{pmatrix}
     c_k^+ & 0 \\ 0 & d_k^+ 
     \end{pmatrix} \ , \quad 
\hat \tau_1^{[k]}  = 
    \begin{pmatrix}
     c_k^- & 0 \\ 0 & d_k^- 
     \end{pmatrix} \ , \quad
\hat \tau_2^{[k]}  = 
    \begin{pmatrix}
    0 & e_k \\ f_k & 0 
     \end{pmatrix} \ , \quad 
\hat \tau_3^{[k]} = \left[\hat \tau_2^{[k]}\right ]^{\text{T}}, 
\end{equation} 
with   
\begin{equation}
\label{cdk}
\begin{dcases}
c_k^{\pm} = a^2\, c_{k-1}^{\pm} + \frac {1}2 \, d_{k-1}^{\pm} \\
d_k^{\pm} = b^2\, c_{k-1}^{\pm} + \frac {1}2 \, d_{k-1}^{\pm}  \ ,
\end{dcases} 
\qquad \qquad
 \begin{dcases}
e_k = \frac a{\sqrt 2} e_{k-1} + \frac b{\sqrt 2} f_{k-1} \\
f_k = \frac b{\sqrt 2} e_{k-1} + \frac a{\sqrt 2} f_{k-1}\ ,
\end{dcases}
\end{equation}
with initial conditions $ c_0^+ = 1, d_0^+ = 0, c_0^- = 0,d_0^- = 1,  e_0 = 1, f_0 = 0$.
Here $a,b$ are the coefficients in Eq. (\ref{eq:state_coeff}) evaluated at the critical point $K_k=K$. 
These recurrence relations admit the exact solution
\begin{equation}
\label{eq:alpha_solution}
 \begin{dcases}
c_k^{+} = \frac {1+2b^2 \, e^{-k\alpha}}{1+2 b ^2} 
\\
d_k^ {+} = 2 b^2  \frac {1-\, e^{-k\alpha}}{1+2 b ^2}
\end{dcases}  
\ , \qquad 
 \begin{dcases}
c_k^{-} = \frac {1-\, e^{-k\alpha}}{1+2 b ^2}\, \\
d_k^{-} = \frac {2  b ^2+e^{-k\alpha}}{1+2 b ^2}
\end{dcases} \ , 
\qquad
 \begin{dcases}
e_k = \frac 12 \left (e^{-\beta k} + e^{-\gamma k}\right ) \\
f_k = \frac 12 \left (e^{-\beta k} - e^{-\gamma k}\right )  \ 
\end{dcases} ,
\end{equation}
with
\begin{equation}
\alpha\equiv \ln \sqrt {K^2+4} \ ,  \qquad \beta \equiv \ln \frac{\sqrt 2}{a-b}, \qquad
\gamma \equiv \ln \frac{\sqrt 2}{a+b} \ .
\end{equation} 
Being the ground state $\ket{\psi}_{\text{GS}}=\ket + ^{[n]}$ as in Eq.  (\ref{eq:gs}),  we just have $\ro_A= \hat \tau^{[k]}_0$, which is already written in diagonal form.
Hence, the exact fixed-point entanglement entropy for this bipartition is 
\be
S_A=- c^+_k \ln c^+_k - d_k^+\ln d_k^+\,.
\ee
Note that, because of the tree structure of the ground state, $S_A$ is a function only of $k$ and not of $k$ and $n$ separately. 
For large $k$ we have the exact asymptotic expansion
\be
\label{eq:exact_pow2}
 S_{A}   = S_* +  c_* e^{-k \alpha} +O(e^{-2k \alpha})  \ ,
\ee
with 
\be
S^* = \ln(1+2 b^2)-\frac{2b^2}{(1+2 b^2)}  \ln 2b^2   , \quad \quad
c^* =  \frac{2b^2\ln 2b^2}{1+2b^2}  . \label{S*}
\ee
Thus, the entanglement entropy saturates to a constant $S^*$, exponentially fast in $k$.
Since $e^{-k \ln2}= (\ell_k/L)$, Eq. (\ref{eq:exact_pow2}) is an expansion for $\ell_k\ll L$.

\begin{figure}[t]
\centering
\fontsize{13}{10}\selectfont
\includegraphics[width=0.62\textwidth]{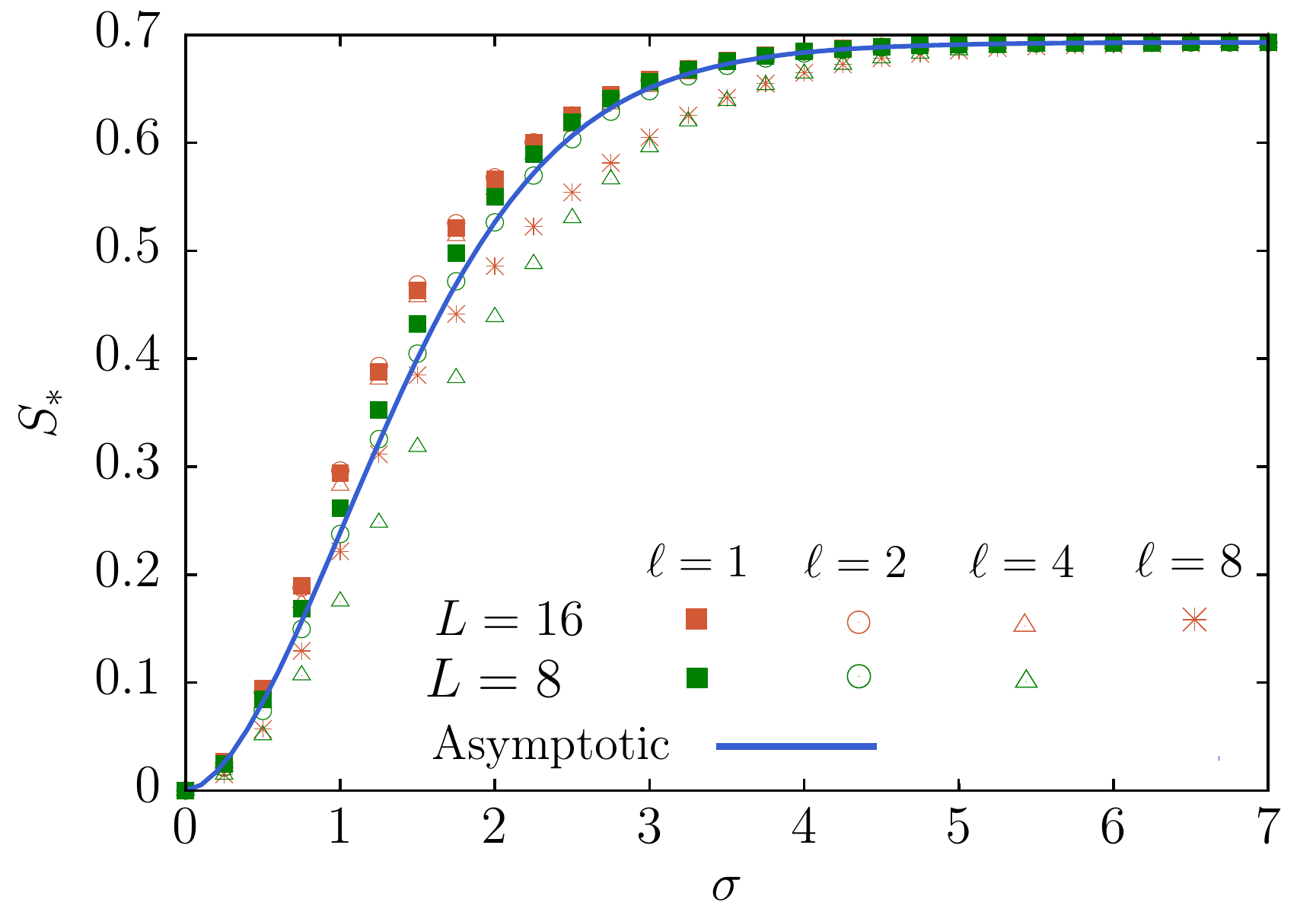}
\caption{Critical entanglement entropy for $\ell_k=2^{n-k}$ for different values of $\sigma$. 
We compare the results obtained by means of exact diagonalisation for $L=8$ (green) and $16$ (red) with $\ell=1,2,4,8$ with the asymptotic saturation value of the 
fixed-point entanglement entropy (\ref{eq:exact_pow2}) (full line). 
Although the systems are very small, the results match quite well for $\ell\ll L$. 
 }
\label{fig:exact_pow2}.
\end{figure}

In Fig. \ref{fig:exact_pow2}, we plot the asymptotic fixed-point value $S_*$ \eqref{S*} as a function of $\sigma$.
We also compare it with the exact diagonalisation results for $L=8$ and $L=16$ with all possible values of $\ell_k=1,2,4$ and $\ell=8$ for $L=16$.
Although the size of the system  is very small, the results for small $\ell$ lie very close to the saturation value 
$S^*$ (expected to be reached for $\ell_k\ll L$, but large $L$).

\subsection{Entanglement entropy for the partition (\ref{Sec:2})}
\label{subsec:one_cut}

\begin{figure}[t]
\centering
\fontsize{13}{10}\selectfont
\includegraphics[width=0.98\textwidth]{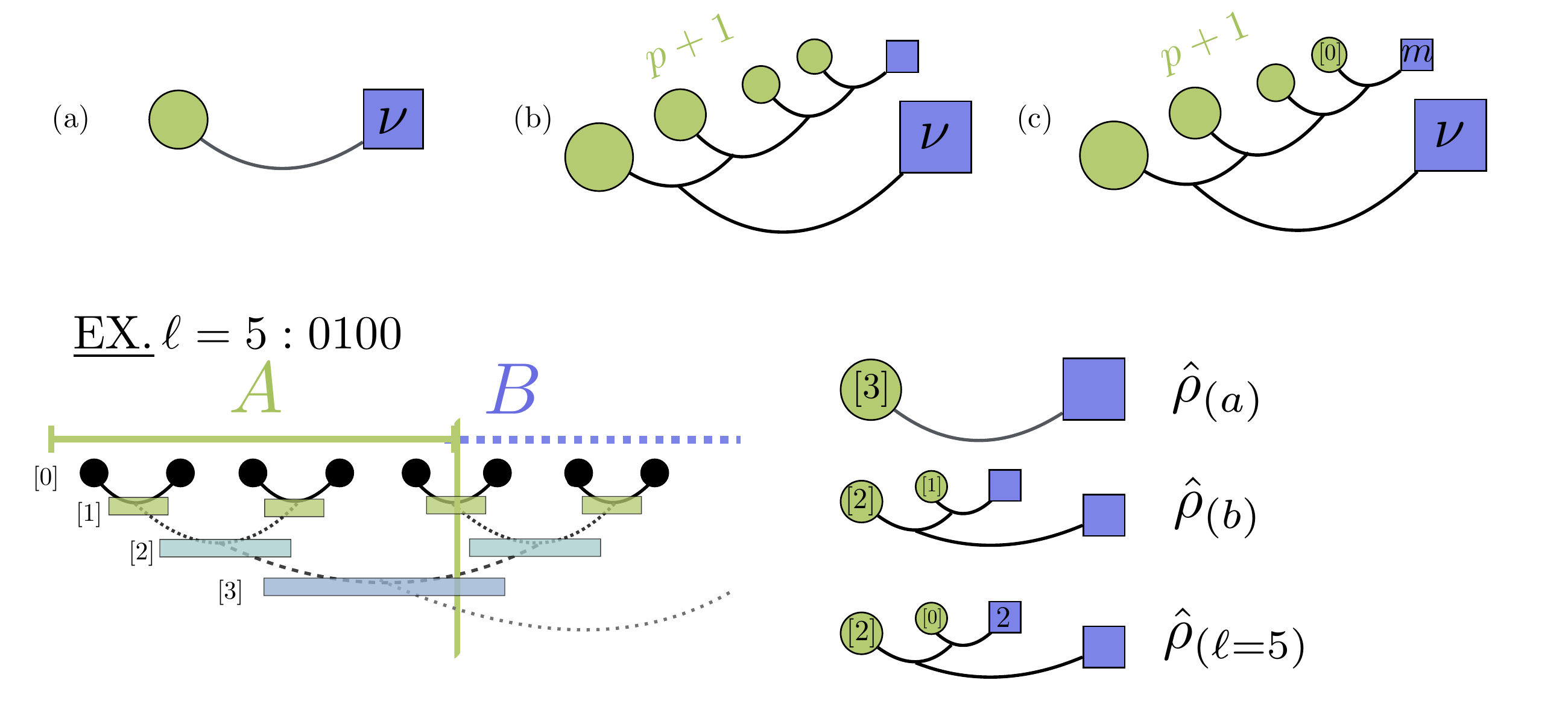}
\caption{Pictorial representation of the construction of the reduced density matrix for the partition (\ref{Sec:2}) of length $\ell_{pm}=(2^p-1)2^m+1$. The green circles represent the basis of the reduced density matrix, while the blue squares are the elements over which the trace has been performed. The label $[k]$ inside the green circles indicates that we refer to the $k$-th effective{ spin $\ket{\pm}^{[k]}$}. 
(Top) General rules: \ref{a} trace away $2^{\nu}$ sites, 
\ref{b} trace away the $(m-1)$-th effective spin on the right, and \ref{c} trace out the remaining $2^m-1$ physical sites from the last effective spin on the right. 
(Bottom) Example for $\ell = 5$ (the value of $L$ is irrelevant) discussed in the main text.}
\label{fig:tree_parti}
\end{figure}

In this subsection, we obtain the renormalised reduced density matrix of the critical ground state (\ref{eq:gs}) for a partition of length $\ell_{pm}=(2^p-1)2^m+1$.
(We can also think to the bipartition as a binary number:  the binary representation of $\ell_{pm}-1$ is just $p$ ones followed by $m$ zeros -- we can place $\nu=n-p-m$ 
zeros in front of it -- the result does not depend on $\nu$).
The number of effective spins broken by the partition is $p+1$, as it should be clear also from Fig. \ref{fig:tree_scaling}.
The rules to construct recursively  the reduced density matrix are:
\begin{enumerate}[label=(\alph*)]
\item Trace away recursively half of the spins $\nu=n-m-p$ times to get the reduced density matrix $\ro_{(a)}$ of the first $2^{p+m}$ spins. 
$\ro_{(a)}$ is a $2 \times 2$ matrix in the basis $\ket{\pm}^{[p+m]}_1$ as in Sec. \ref{sec:power2}. 
\label{a}
\item Trace out the $(m-1)$-th effective spin on the right.  $\ro_{(b)}$ is a $2^{p+1} \times 2^{p+1}$ matrix in the 
basis $\ket{\pm}^{[m+p-1]}_1\otimes\ket{\pm}^{[m+p-2]}_3\otimes\dots\otimes \ket{\pm}^{[m]}_{2^p-1}\otimes \ket{\pm}^{[m-1]}_{2^{p+1}-1}$.
This is the reduced density matrix of the first $2^{m+p}-2^{m-1}$ spins at level zero. 
\label{b}
\item On the rightmost effective spin at level $m-1$, trace out all spins except one at level zero by means of the recurrence relations  \eqref{eq:alpha_def}. 
This last step does not increase the size of the reduced density matrix because all involved effective spins have been already broken before. 
$\ro_{(c)}$ is the desired  $2^{p+1} \times 2^{p+1}$ density matrix in the 
basis $\ket{\pm}^{[m+p-1]}_1\otimes\ket{\pm}^{[m+p-2]}_3\otimes\dots\otimes \ket{\pm}^{[m]}_{2^p-1}\otimes \ket{\pm}^{[0]}_{\ell_{pm}}$.
\label{c}
\end{enumerate}
These three rules are pictorially summarised in Fig. \ref{fig:tree_parti}. 

As an example, let us consider the case $p=1$ and $m=2$, i.e. an interval of length $\ell=5$ (in a chain of arbitrary size $L$): 
$\ell-1=4\to \dots 0100$, see also Fig. \ref{fig:tree_parti} (bottom). 
\ref{a} Compute first the $2\times2$ reduced density matrix of the first spin at level $p+m=3$, i.e. $\ro_{(a)} = a^2\, \ket+\bra+^{[3]}_1 + b^2 \ket-\bra-^{[3]}_1$ . 
\ref{b} Trace away the rightmost effective block-spin at level one.  
This yields the $4\times 4$ matrix $\ro_{(b)}$, in the basis $\ket{\pm}^{[2]}_1\otimes \ket{\pm}^{[1]}_3$. 
\ref{c} Trace out the sixth physical spin  from the block spin at level one.
Hence $\ro_{\ell=5}$ is then a $4 \times 4$ matrix in the {basis $\ket{\pm}^{[2]}_1\otimes \ket{\pm}^{[0]}_5$}. 

Step \ref{a} has been already discussed in Sec. \ref{sec:power2}. For the ground state (\ref{eq:gs}), Eqs. (\ref{eq:alpha_def}-\ref{eq:alpha_basis}) evaluated at $k=\nu$ read
\begin{equation} \label{eq:qui}
\ro_{(a)} =
    c_{\nu}^+\,  \ket +\bra + _1^{[p+m]} + d_{\nu}^+\, \ket - \bra - _1^{[p+m]} \ ,
\end{equation}
where $c_{\nu}^{\pm}$ are given in Eq. (\ref{eq:alpha_solution}).  The reduced density matrix is a $2 \times 2$ matrix in the basis of the effective spin at level $n-\nu = p+m$.

Now we perform step \ref{b} and construct $\hat \rho_{{(b)}}$. 
We trace away the rightmost $(m-1)$-th effective spins from the reduced density matrix in Eq. (\ref{eq:qui}). 
Since the partial trace occurs at level $m$, it leaves untouched all the $p$ effective spins on the left. 
The trace of the $(m-1)$-th right effective spin from the $m$-th yields
\bea
\fl \label{eq:properties_wowo_3}
 \hat \rho^{[0]}_0 \equiv \tr_{m-1}\, \ket + \bra+_1^{[m]} &=& \rozero \ ,
 \quad  
 \hat \rho^{[0]}_1 \equiv \tr_{m-1}\, \ket - \bra-_1^{[m]}=\rouno \ , 
\nn\\ \fl
 \hat \rho_2^{[0]}\equiv \tr_{m-1}\, \ket + \bra-_1^{[m]} &=&\rodue\ , \quad  
 \hat\rho^{[0]}_3 \equiv \tr_{m-1}\, \ket - \bra+_1^{[m]}= \big( \ro_2^{[0]} \big)^T .
 \eea
Let us now define for convenience the following matrices for a generic  $p$
\bea
& \ro^{[p]}_0  \equiv    \tr_{m-1}\, \ket + \bra+_1^{[p+m]}, \quad \quad 
 \ro^{[p]}_1  \equiv    \tr_{m-1}\, \ket - \bra-_1^{[p+m]}  ,\nn \\
& \ro^{[p]}_2  \equiv    \tr_{m-1}\, \ket +\bra-_1^{[p+m]} ,\quad \quad
 \ro^{[p]}_3  \equiv    \tr_{m-1}\, \ket -\bra+_1^{[p+m]} \ .
\eea
They are obtained recursively as
\begin{align}
\label{eq:recurrence_matrix}
& \ro^{[p]}_0  
    =
    a^2 \sigmaup \otimes \ro^{[p-1]}_0
    + b^2 \sigmadown \otimes \ro^{[p-1]}_1
    + a\, b \sigmaplus \otimes \ro^{[p-1]}_2 + h.c. \ , 
     \nn  \\ 
& \ro^{[p]}_1  
    =
    \frac 12 \sigmaup \otimes \ro^{[p-1]}_1
    + \frac 12 \sigmadown \otimes \ro^{[p-1]}_0
    + \frac 12 \sigmaplus \otimes \ro^{[p-1]}_3 + h.c.\ ,
\\
& \ro^{[p]}_2    =\ro_3^{[p]\, T}=
    \frac a{\sqrt 2} \sigmaup \otimes \ro^{[p-1]}_2
    +  \frac a{\sqrt 2} \sigmaplus \otimes \ro^{[p-1]}_0  
    +  \frac b{\sqrt 2} \sigmadown \otimes \ro^{[p-1]}_3 \nn\\ &\qquad\qquad\qquad
    +  \frac b{\sqrt 2} \sigmaminus \otimes \ro^{[p-1]}_1\ , \nn
    \end{align}
with the initial conditions given by Eq. (\ref{eq:properties_wowo_3}). At the end of the recursive construction, the reduced density matrix 
is a $2^{p+1}\times2^{p+1}$ block-diagonal matrix with
symmetric and anti-symmetric components
\begin{equation}
\label{eq:rho_p1}
\ro_{(b)} = \hat \rho_{l_{p1}} = c_{\nu}^+\,  \ro^{[p]}_0  + d_{\nu}^+\, \ro^{[p]}_1 
    = {
    \begin{pmatrix}
    \ro_S^{[p]} & 0\\
    0 &    \ro_A^{[p]}
    \end{pmatrix}}\ ,
\end{equation}
as it follows from the fact that the Hamiltonian  (\ref{h0dyson}) is block-diagonal. The $\ro_{S/A}^{[p]}$ are $2^p\times 2^p$ matrices given directly in terms of Eqs. (\ref{eq:alpha_solution}-\ref{eq:recurrence_matrix}). 

We are ready for the final step \ref{c}.
The reduced density matrix $\ro_{(c)}$ is obtained by tracing away the remaining $2^m-1$ physical spins from  (\ref{eq:rho_p1}). This is achieved by means of the procedure (\ref{sec:0}) applied to the $m$-th effective spin $\ket{\pm}^{[m]}_{2^{p+1}-1}$.  
The final reduced density matrix then reads
\begin{equation}
\label{eq:rho_pm}
\hat \rho_{(c)} = c_{\nu}^+\, {\ro}^{[p,m]}_0  + d_{\nu}^+\, \ro^{[p,m]}_1 \ ,
\end{equation}
where ${\ro}^{[p,m]}_i$ are given by the recurrence relations (\ref{eq:recurrence_matrix}) with
 initial conditions given by ${\ro}^{[0,m]}_i=\hat \tau^{[m]}_i$ in Eq. (\ref{eq:alpha_basis}). 

\begin{figure}[t]
\centering
\fontsize{13}{10}\selectfont
\includegraphics[width=0.98\textwidth]{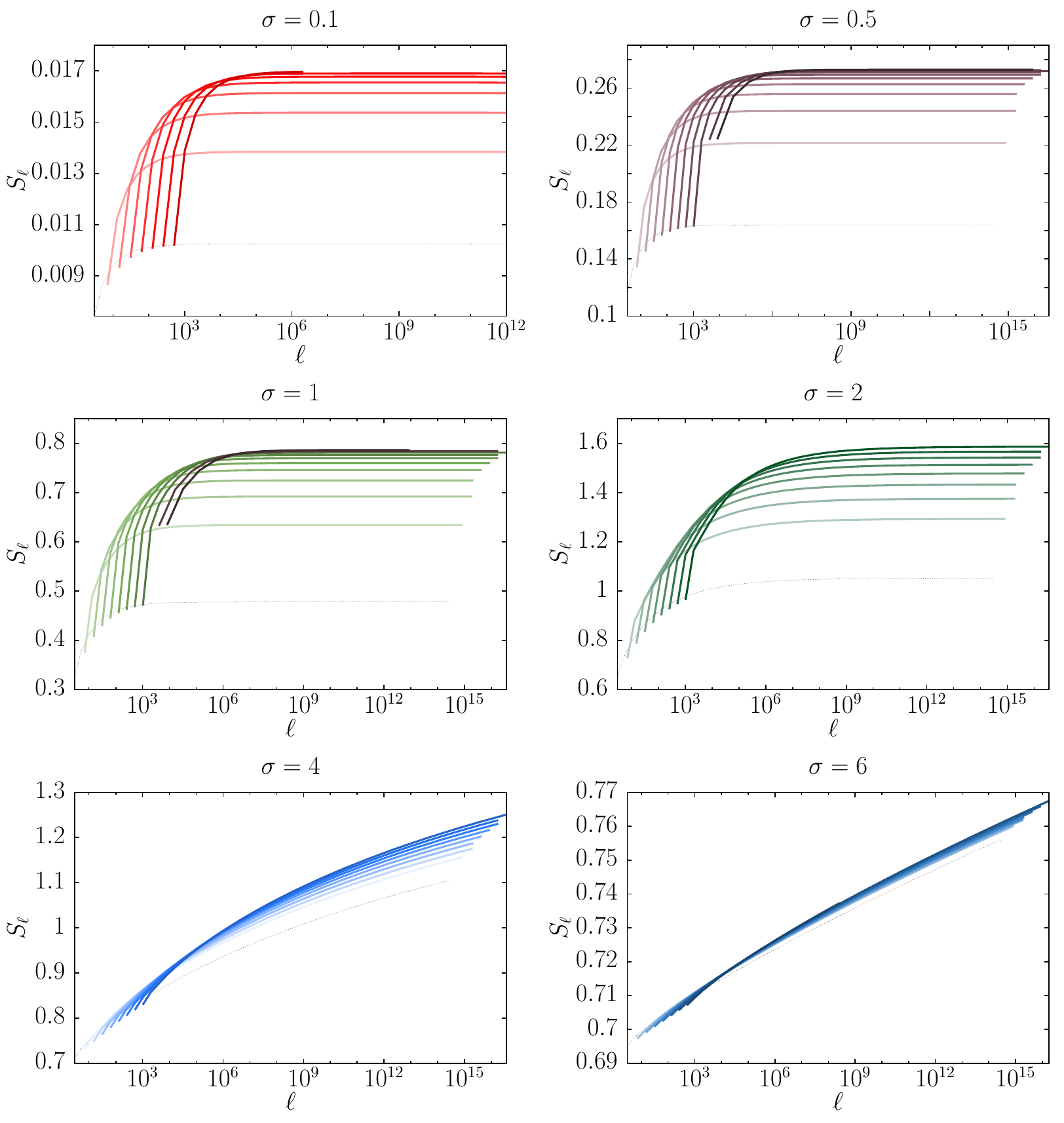}
\caption{Fixed-point entanglement entropy for the partition (\ref{Sec:2}). 
Each curve in the plot corresponds to a fixed value of $p$: $\ell_{pm}$ is changed varying $m$. 
Different curves refers to increasing values of $p=1, \dots ,9$. 
The dotted black lines correspond to $p=1$.
}
\label{fig:varying_p}
\end{figure}

At this point, we have a general recursive form for the reduced density matrix whose elements can be easily constructed. 
We diagonalise  the reduced density matrix (\ref{eq:rho_pm}) to yield the entanglement entropy.
The results of this procedure are shown in Fig. \ref{fig:varying_p} for various values of $\sigma$. 
We plot the entanglement entropy as function of $\ell$ for several values of $p$ and $m$ (as we stressed, the value of $L$ does not matter at the fixed point).
For fixed $p$, $S_{A}$ initially grows with $m$ up to a saturation value depending on $p$. 
The same trend holds when $p$ is varied at fixed $m$. 
The scaling with $\ell$ approaching its asymptotic value depends on $\sigma$. 
In fact, for small $\sigma $ the entanglement entropy initially grows quickly and saturates after a given $\ell^*(\sigma)$.
This saturation  value $\ell^*(\sigma)$ increases  as  $\sigma$ gets larger.
Conversely,  at  large $\sigma$, we observe a very small growth, roughly compatible with a logarithmic behaviour that persists for many decades (more than 15) before saturation.
We will get an analytic understanding of this behaviour from the expansion at large $\sigma$.
For fixed $\ell$ and $L$, the dependence of the entanglement entropy on $\sigma$ is very similar to the one in Fig. \ref{fig:one_cutEE_sigma}: starting from zero, 
it grows with $\sigma$ up to a maximum and then decreases.

\subsubsection{Some analytical expansions.}

We derive an analytical expression of the entanglement entropy in the case $p=1$, which corresponds to an interval of length $\ell_m= 2^m+1$. 
In this case, the reduced density matrix is just a $4\times4$ density matrix in the basis $\ket{\pm}^{[m]}_1\otimes\ket{\pm}^{[0]}_{\ell_m}$, 
whose elements are determined recursively via Eq. (\ref{eq:rho_pm}). 
These elements are easily worked out analytically and so are their expansion for  $\sigma\ll1$ and $\sigma\gg 1$. 
See \ref{app:p1proof} for the derivation and the details.

\begin{figure}[t]
\centering
\fontsize{13}{10}\selectfont
\includegraphics[width=0.33\textwidth]{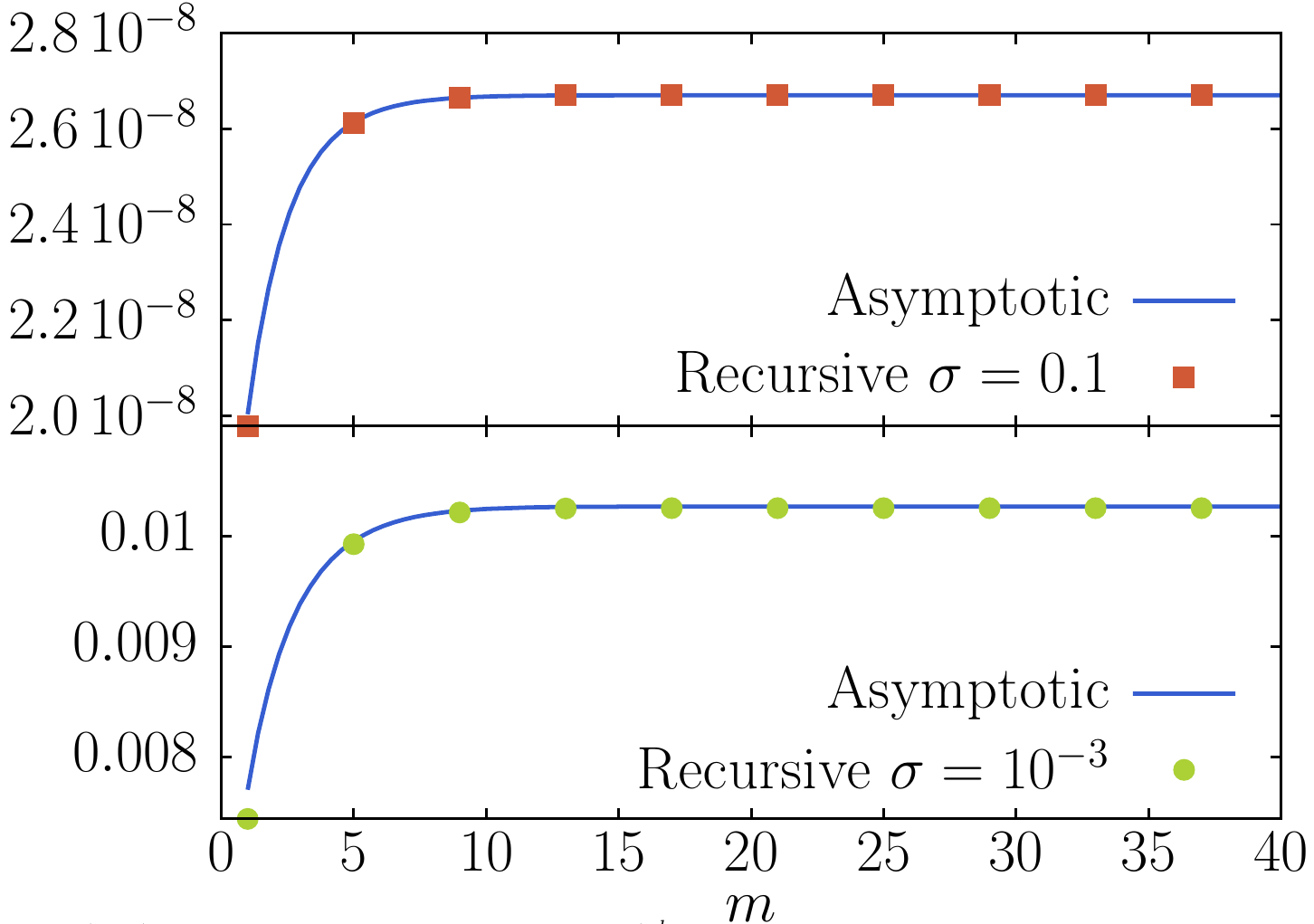}
\includegraphics[width=0.33\textwidth]{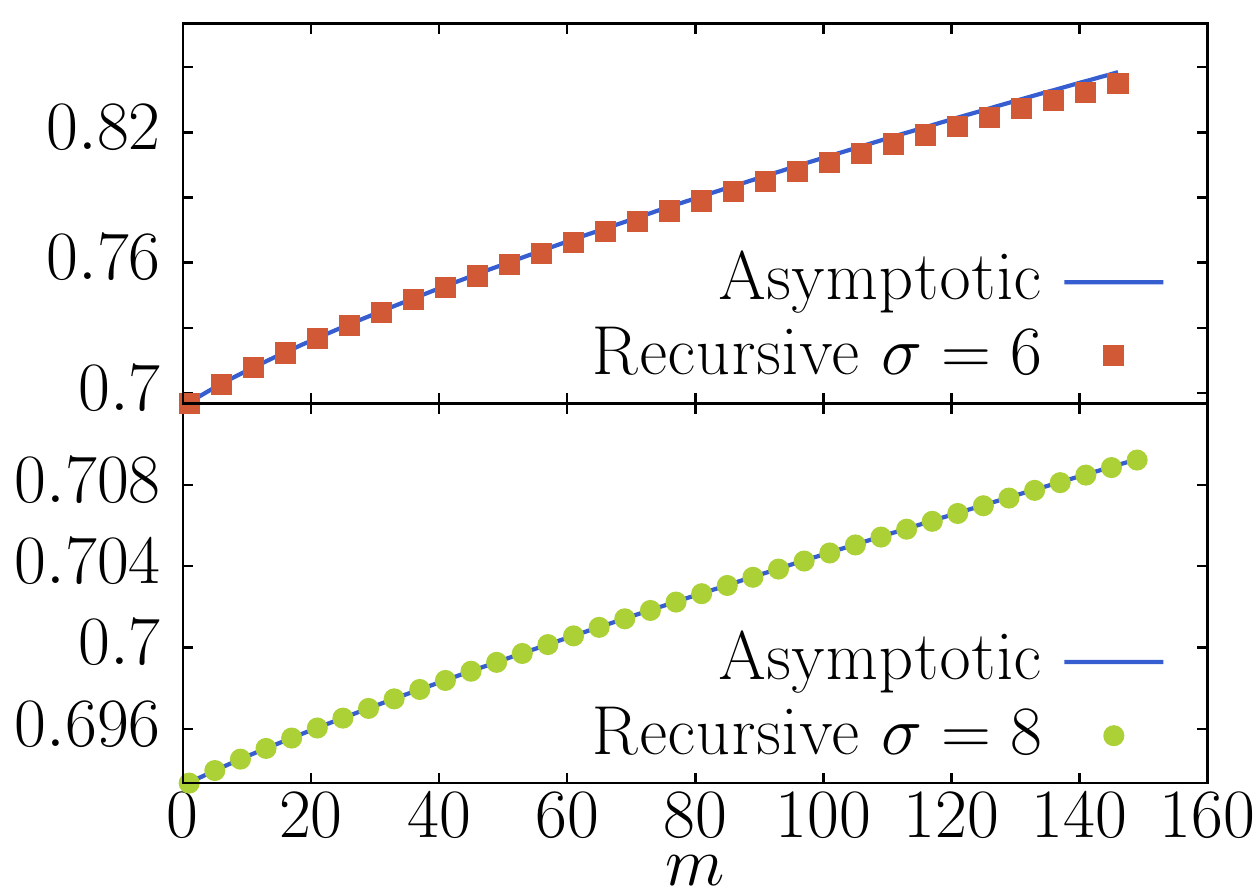}
\includegraphics[width=0.31\textwidth]{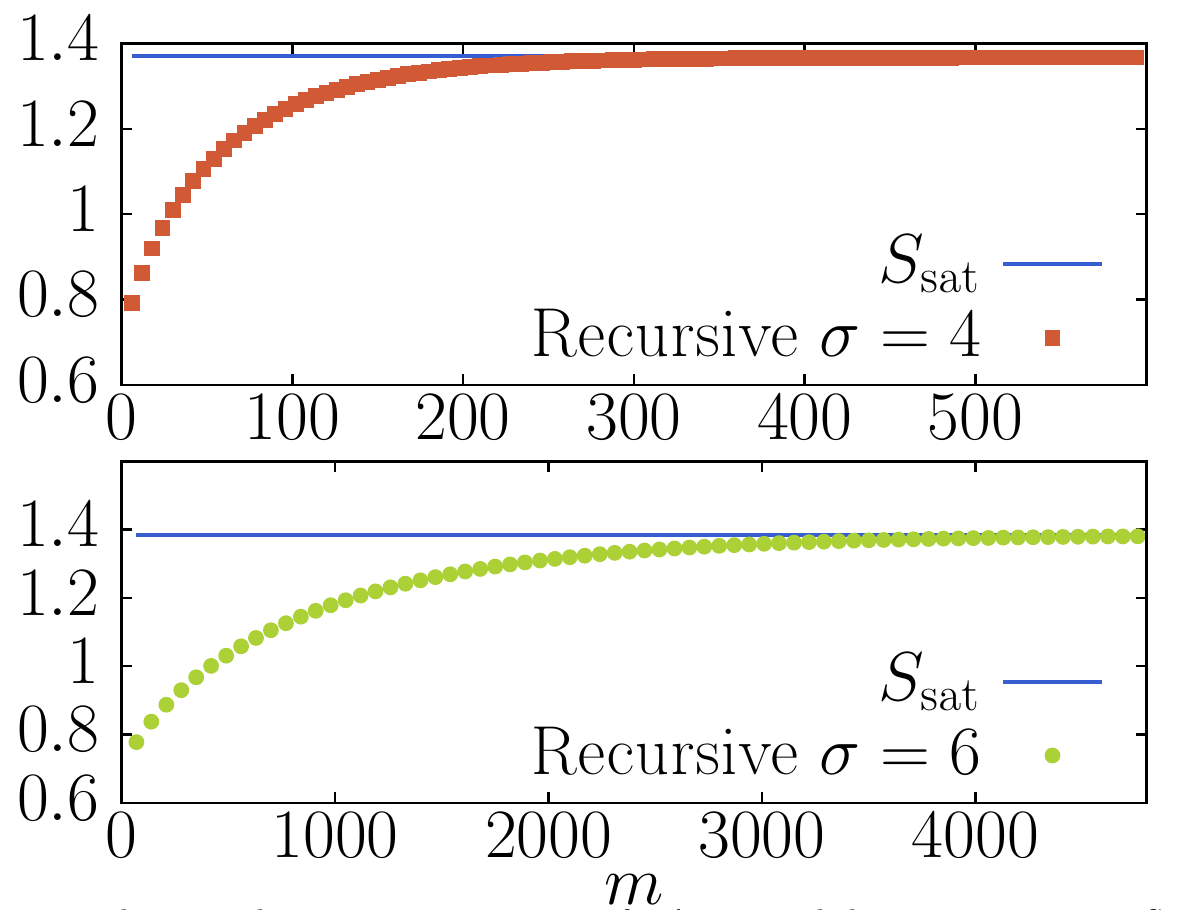}
\caption{Entanglement entropy at the critical point for the partition (\ref{Sec:2}) with $\ell_{m}=2^m+1$. 
We compare the asymptotic expressions with the entanglement entropy obtained from the recursion relations (\ref{eq:rho_pm}). 
(Left) For  $\sigma \ll 1$, the entanglement saturates quickly to a constant value in Eq. (\ref{eq:exact_smallsigma}).  
(Center) For $\sigma \gg 1$ and for $m\ll K^2$, tit grows logarithmically with $\ell_m$ as in Eq. (\ref{eq:exact_bigsigma}). 
(Left) For $\sigma \gg 1$ and for $m\gg K^2$, it saturates to \eqref{Ssat}.
}
\label{fig:exact_results}
\end{figure}

For $\sigma \ll 1$, the critical coupling $K\to 0$, the expansion of the matrix elements (\ref{eq:rho_pm}) at order $\mathcal O(K^3)$ provides the following asymptotic expression
for large $\ell_m$
\begin{multline}
\label{eq:exact_smallsigma}
 S_{\ell_m}  =  \frac {K^2}{4} \left
    (1+\ln \frac 8{K^2} \right)\,
    -\frac {K^2}{16} \frac 1{\ell_m-1}
      \left (\ln \frac {8}{K^2} +\frac 14 \left [\log_2(\ell_m-1)+2\right ]^2
      \right ) 
      + \mathcal O(K^3 ) \ .
\end{multline}
Here we first take the limit of small $K$ and only after large $\ell$: the two limits do not commute.  
The entanglement entropy saturates for large $\ell$ to a constant value proportional to $K^2\ln K$. 
On the left panels of Fig. \ref{fig:exact_results}, we compare the entanglement entropy in Eq. (\ref{eq:exact_smallsigma}) with the one from the 
recurrence relations (\ref{eq:rho_pm}), finding a perfect match of the results. 

In the opposite limit $\sigma \gg 1$, i.e. for critical coupling $K\to \infty$, we find two different regimes for large $\ell_m$.
At fixed  $\ell_m$, for $K^{-1}\ll \ell_m\ll 2^{K^2}$, the entanglement entropy has the following asymptotic expansion
\be \label{eq:exact_bigsigma}
 S_{\ell_m}  = \ln 2 +\frac{\log_2(\ell_m-1)}{4K^2} \Big\{ 1 -   \ln \Big[\frac{\log_2(\ell_m-1)}{4 K^2} \Big ]
     \Big \} + \mathcal O\left (K^{-3}, (\log_2 \ell_m)^2K^{-4}\right) \ .
\ee
In this regime, the entanglement entropy grows logarithmically with the length of the interval, but with a very small prefactor proportional to $K^{-2}$, 
as it was clear already from the data in Fig. \ref{fig:varying_p}. 
This prediction is checked against the exact data from the recurrence relations in Fig. \ref{fig:exact_results} (right panels), finding perfect agreement.
Furthermore, the results  in Fig. \ref{fig:varying_p}  for $\sigma \gg 1$ show that the same logarithmic growth, numerically with the same prefactor, 
appears for all other lengths with $p\neq1$, although we do not have an analytic handle on them. 
It is worth mentioning that this behaviour is reminiscent of the standard one for critical short-range systems \cite{cc-04}.
However, the right panels of Fig. \ref{fig:exact_results}, show also that the exact data start deviating from this logarithmic scaling for  $\ell_m\sim 2^{K^2}$. 
Indeed, at this value of $\ell_m$, saturation toward the asymptotic value starts taking place and  Eq. (\ref{eq:exact_bigsigma}) fails: 
in other words the limits $K\to\infty$ and $m\to \infty$ do not commute. 
In turn, when $\ell_m \gg 2^{K^2}$, the entanglement entropy saturates to 
\begin{equation}
S_{\text{sat}} = 2 \ln 2 -\frac 1{K^2} + \mathcal O(K^{-3}) \ .
\label{Ssat}
\end{equation}

\subsection{Entanglement entropy for the partition (\ref{sec:1})} 
\label{subsec:two_cut}
We now consider the partition (\ref{sec:1}) as in Fig. \ref{fig:tree_scaling} (bottom). 
This partition cuts $n$ effective spins, giving rise to a $L\times L$ reduced density matrix. 
 The ground-state density matrix can be written as a function of the $(n-1)$-th effective spins as in Eq. (\ref{eq:gs_recu})
\begin{multline}\label{eq:rho_gs}
\hat \rho_{\text{GS}} = a^2\,  \ket {+} \bra {+}_1^{[n-1]} \otimes  \ket {+} \bra {+}_2^{[n-1]} \, +
    b^2 \, \ket {-} \bra {-}_1^{[n-1]} \otimes  \ket {-} \bra {-}_2^{[n-1]} \\ +
    a b\,  \ket {+} \bra {-}_1^{[n-1]} \otimes  \ket {+} \bra {-}_2^{[n-1]} +  a b\,  \ket {-} \bra {+}_1^{[n-1]} \otimes  \ket {-} \bra {+}_2^{[n-1]} \ ,
\end{multline}
where the coefficients $a, b$ are given by Eq. (\ref{eq:state_coeff})
We trace away the sub-system $A$ and write the density matrix in the basis represented pictorially by the green circles in Fig. \ref{fig:tree_scaling} (bottom-right).
%
This trace over $A$ is relatively simple, because it can be performed separately on the first and on the second $(n-1)$-th effective spin, see Fig. \ref{fig:tree_scaling}. 
On the left, the partial trace is equivalent to the procedure (\ref{sec:0}) applied to $\ket{\pm}\bra{\pm}_1^{[n-1]}$. It results in the $2\times 2$ matrices ${\hat \tau_i^{[n-1]}}$ defined in Eq. (\ref{eq:alpha_def}) in the basis $\ket{\pm}^{[0]}_1$ of the first physical spin.  
On the second block spin, the result obtained from the trace corresponds to Eq. (\ref{eq:recurrence_matrix}) with $p=n-2$ in the basis of 
$\ket{\pm}^{[0]}_{2^{n-1}+2}\otimes \ket{\pm}^{[1]}_{2^{n-2}+2}\otimes \dots \ket{\pm}^{[n-2]}_4$. 
Accordingly, the reduced density matrix is 
\begin{equation}\label{eq:rho_part1_long}
\hat \rho_{A} = a^2\, \hat \tau^{[n-1]}_0 \otimes \hat \rho^{[n-2]}_0 \, +
    b^2 \, \hat \tau^{[n-1]}_1 \otimes \hat \rho^{[n-2]}_1 +
    a b\,  \hat \tau^{[n-1]}_2 \otimes \hat \rho^{[n-2]}_2 +  a b\,  \hat \tau^{[n-1]}_3 \otimes \hat \rho^{[n-2]}_3 \ .
\end{equation}

\begin{figure}[t]
\centering
\fontsize{13}{10}\selectfont
\includegraphics[width=0.48\textwidth]{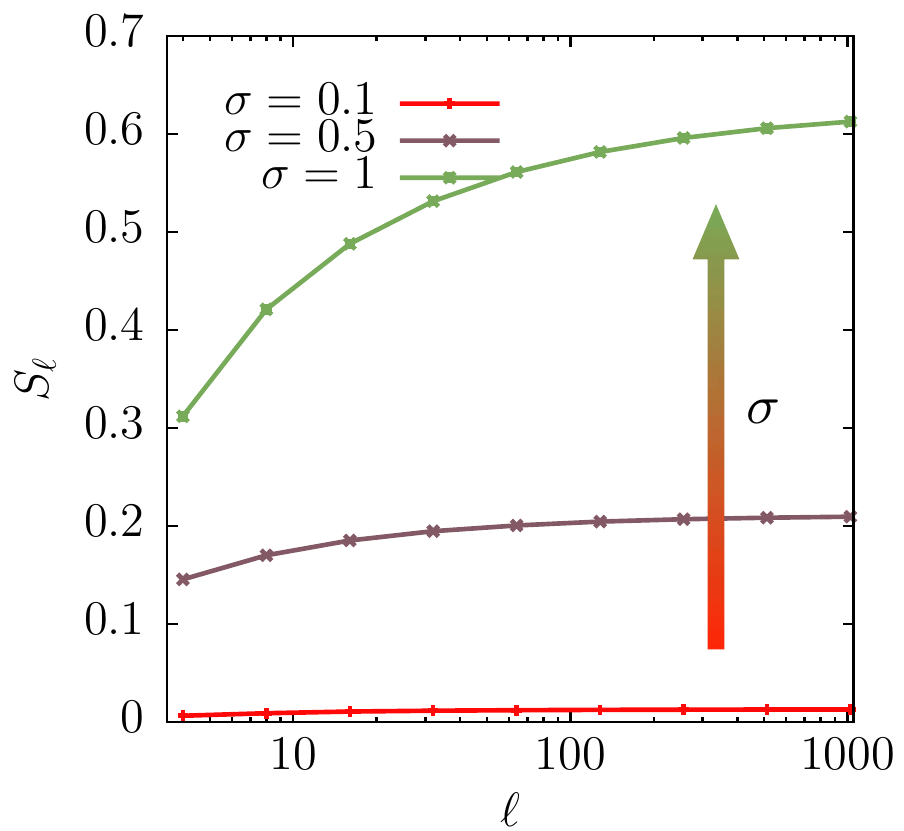}
\includegraphics[width=0.48\textwidth]{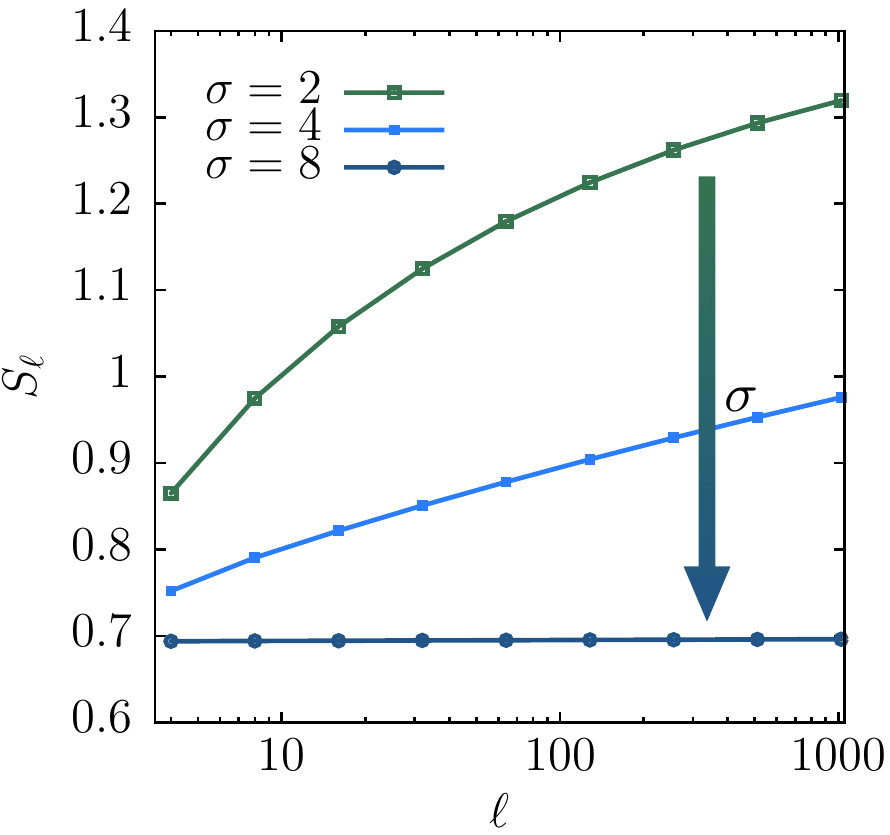}
\caption{Entanglement entropy $S_{\ell}$ with the subsystem length $\ell$ at the critical point with the partition (\ref{sec:1}). (Left) Results for $\sigma=0.1,0.5,1$. (Right) $\sigma=2,4,8$.  For $\sigma \gg 1$, $S_l\sim \ln \ell$ as for the partition (\ref{Sec:2}), as in Fig. \ref{fig:exact_results}. }
\label{fig_boudaries_compa}
\end{figure}

The resulting entanglement entropy reported in Fig. \ref{fig_boudaries_compa} as a function of $\ell=L/2$
has the same qualitative behaviour of the  partition (\ref{Sec:2}): 
it initially grows with the system size and then saturates to a finite value. 

\begin{figure}[t]
\centering
\fontsize{13}{10}\selectfont
\includegraphics[width=0.48\textwidth]{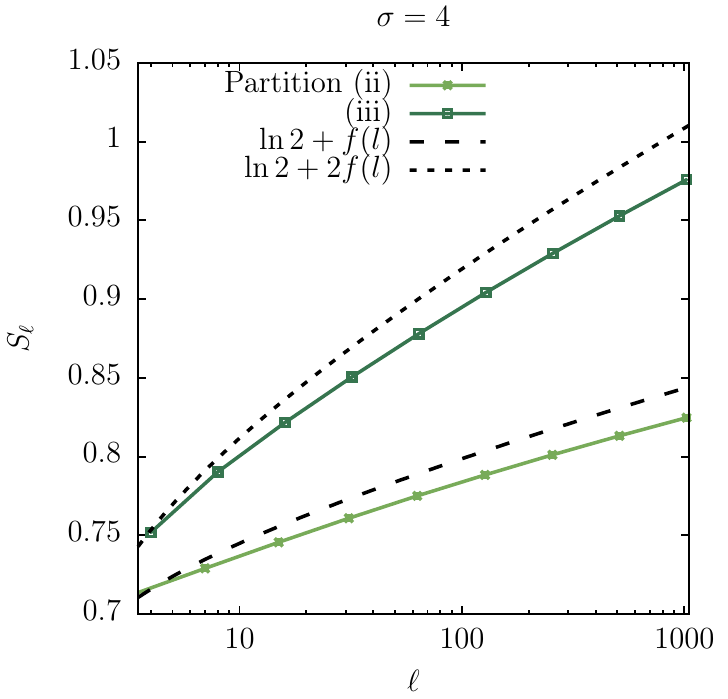}
\includegraphics[width=0.49\textwidth]{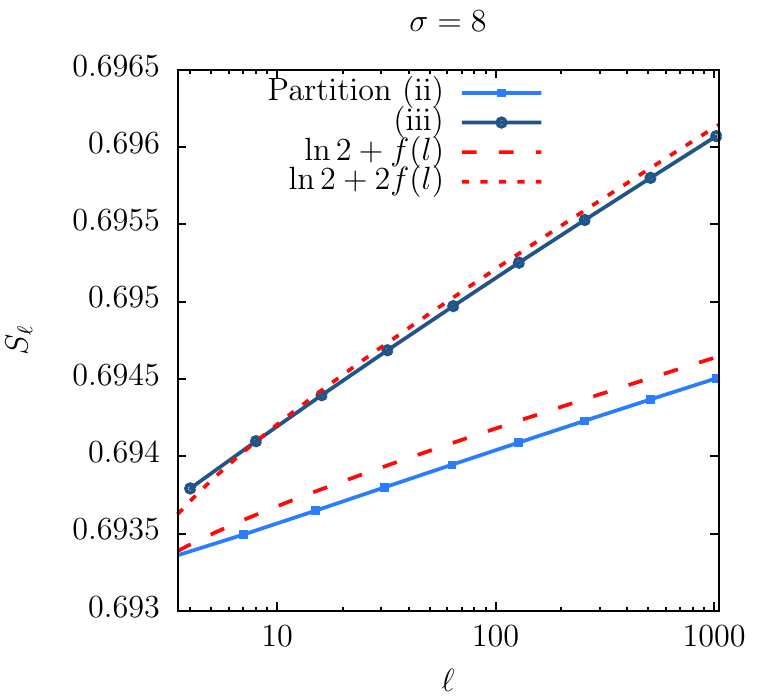}
\caption{Comparison between the entanglement entropy $S_{\ell}$ for the partitions (\ref{Sec:2}) and (\ref{sec:1}), respectively of length $\ell=L/2-1$ and $\ell=L/2$, 
at the critical point for $\sigma \gg 1$. (Left) $\sigma = 4$. (Right) $\sigma = 8$.
The full lines are the entanglement entropies obtained from the recursive density matrices (\ref{eq:rho_pm}) for $\ell=L/2-1$ ($m=1, p=n-2$) 
and (\ref{eq:rho_part1_long}) for $\ell=L/2$.
The dashed lines are Eq.  (\ref{eq:slfl}).
 }
\label{fig_boudaries_compa_bello}
\end{figure}

For large $\sigma$ we can make also a more quantitive comparison.
We focus on the partition  (\ref{Sec:2}) with $m=1$ and $p=n-2$ in (\ref{eq:rho_pm}), i.e. $\ell=L/2-1$ which is the choice that scales more similar 
to the partition of length $\ell=L/2$ considered in this subsection.
These two entanglement entropy are compared in Fig. \ref{fig_boudaries_compa_bello}.
We have analytic result for partition (\ref{Sec:2}) only for $p=1$ and in the regime $K^{-1} \ll \ell\ll 2^{K^2}$, 
when
\begin{equation}\label{eq:slfl}
S_{\ell}   = \ln 2 +  f(\ell)\ .
\end{equation}
with $f(y+1) = (1/4K^2\ln 2) \ln y\left[1-\ln(\ln y/4K^2\ln 2) \right ]$).
We have noticed in Fig. \ref{fig:varying_p} that the same growth with $f(\ell)$ is approximately valid for all values of $p$, and indeed 
also in Fig. \ref{fig_boudaries_compa_bello} the data for $m=1$ and $p=n-2$ are well described by Eq. \eqref{eq:slfl}.
For partition (\ref{sec:1}), the same form describes very well the data, just multiplying $f(\ell)$ by 2. 
This factor is likely  related to the number of boundaries in the partitions, in analogy to what happens for systems with short-range interaction, 
also at the critical point \cite{cc-04,cc-09}. 

\begin{figure}[t]
\centering
\fontsize{13}{10}\selectfont
\includegraphics[width=0.49\textwidth]{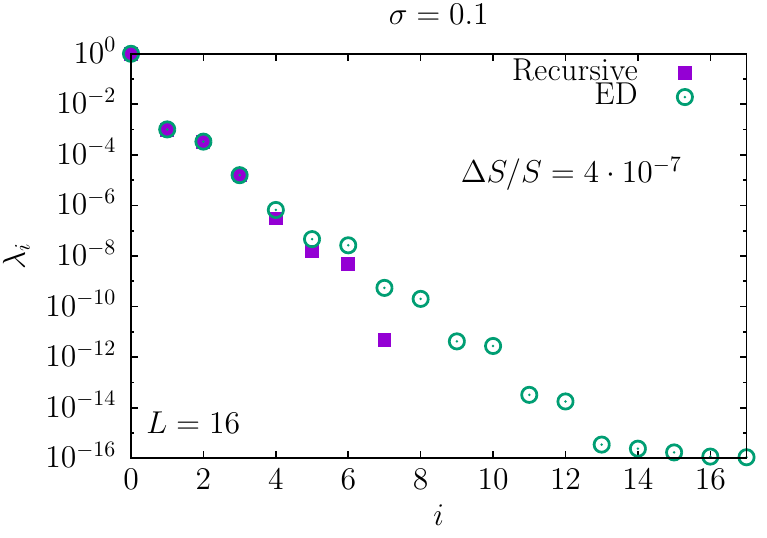}
\includegraphics[width=0.49\textwidth]{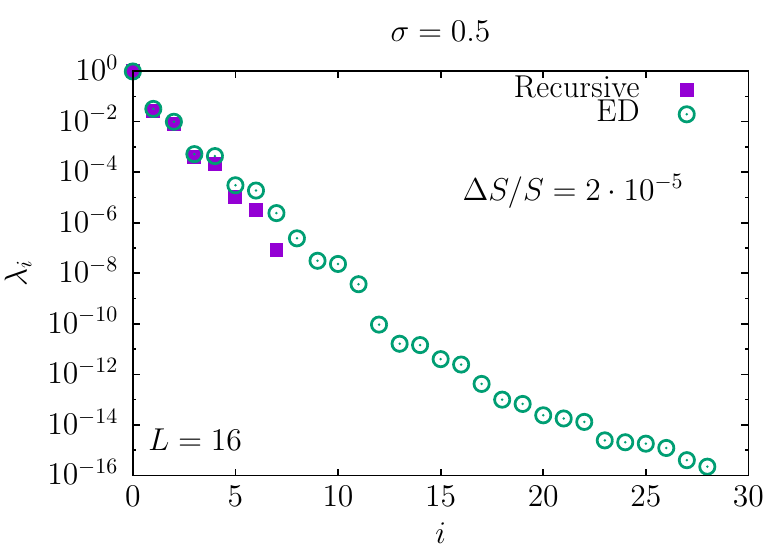}
\includegraphics[width=0.49\textwidth]{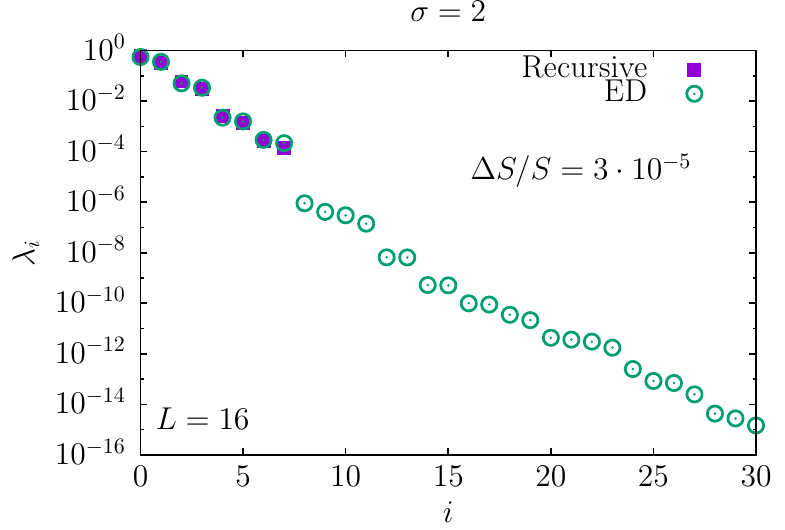}
\includegraphics[width=0.49\textwidth]{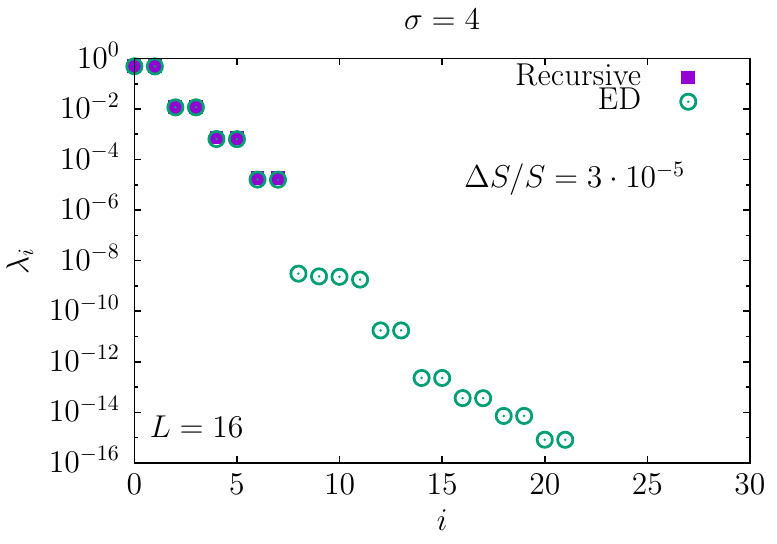}
\caption{Comparison between the entanglement spectrum $\{\lambda_i\}$ obtained by exact diagonalisation and the recursive projection for the partition (\ref{sec:1}). 
The data are for $L=16$. The agreement for the largest eigenvalues is excellent, although the 
fixed-point RG reduced density matrix has $\text{Rank}[\ro_{A}]=8$ and cannot capture the small one. 
The relative difference between the exact and RG entanglement entropy is at most of order $\Delta S/S\leq 10^{-5}$. }
\label{fig:caldervslmp}
\end{figure}

\subsection{Finite rank reduced density matrix.}
\label{sec:general_beha}

We show now by a very elementary argument that the rank of the reduced density matrix is finite.
We focus here on the partition (\ref{sec:1}), but the same reasoning applies to the partition (\ref{Sec:2}) and also others that we did not consider. 
At criticality the reduced density matrix of partition (\ref{sec:1}) can always be written from Eq. (\ref{eq:rho_part1_long}) as
\begin{equation}\label{eq:rho_part1}
\hat \rho_{A} = \sum_{i=0}^3\, c_i\, \hat \tau^{[n-1]}_i \otimes \hat \rho^{[n-2]}_i \ , 
\end{equation}
where $\vec {c} = (a^2,b^2, ab, ab) $, $\hat \tau^{[n-1]}_i$ are $2\times2$ matrices, whose coefficients have been determined exactly (cf. Eqs. (\ref{eq:alpha_basis}-\ref{eq:alpha_solution})), and $\hat \rho^{[n]}_i$ are defined recursively in Eq. (\ref{eq:recurrence_matrix}). 
The structure of $\hat \rho^{[n]}_i$ is such that for all $n$ 
\begin{equation}\label{eq:rank1}
\Rank\left[\rho_i^{[n]}\right]=2 \ .
\end{equation}
From the sub-additivity of the rank it follows that $\text{Rank}[\hat \rho_{A}]\leq 16$. 
For partition (\ref{sec:1}) this bound can be improved to $\text{Rank}[\hat \rho_{A}]= 8$, but for other partitions is tight. 
In \ref{app:finite_redu}, we provide a proof of Eq. (\ref{eq:rank1}) based on linear algebra. An analogous proof might be obtained using tensor networks. 

The real-space RG procedure projects the ground state onto a finite entanglement state. 
In Fig. \ref{fig:caldervslmp}, we compare the spectrum of the reduced density matrix -aka the \emph{entanglement spectrum}- of the fixed-point RG ground state (\ref{eq:rank1}) 
to the one computed from the ground state of from exact diagonalisation. 
In the RG procedure, we have only eight non-zero eigenvalues. 
These match extremely well the largest eigenvalues of the exact reduced density matrix.
 Consequently,  the relative difference between the exact entanglement entropy and the one computed with eight eigenvalues is small for all values of $\sigma$ and 
 it is at most of order $\Delta S/S  \leq 10^{-5}$. 
This suggests that the portion of the entanglement spectrum not captured by the RG procedure is irrelevant for the exact entanglement entropy in the thermodynamic limit.

\begin{figure}[t]
\centering
\includegraphics[width=0.7\textwidth]{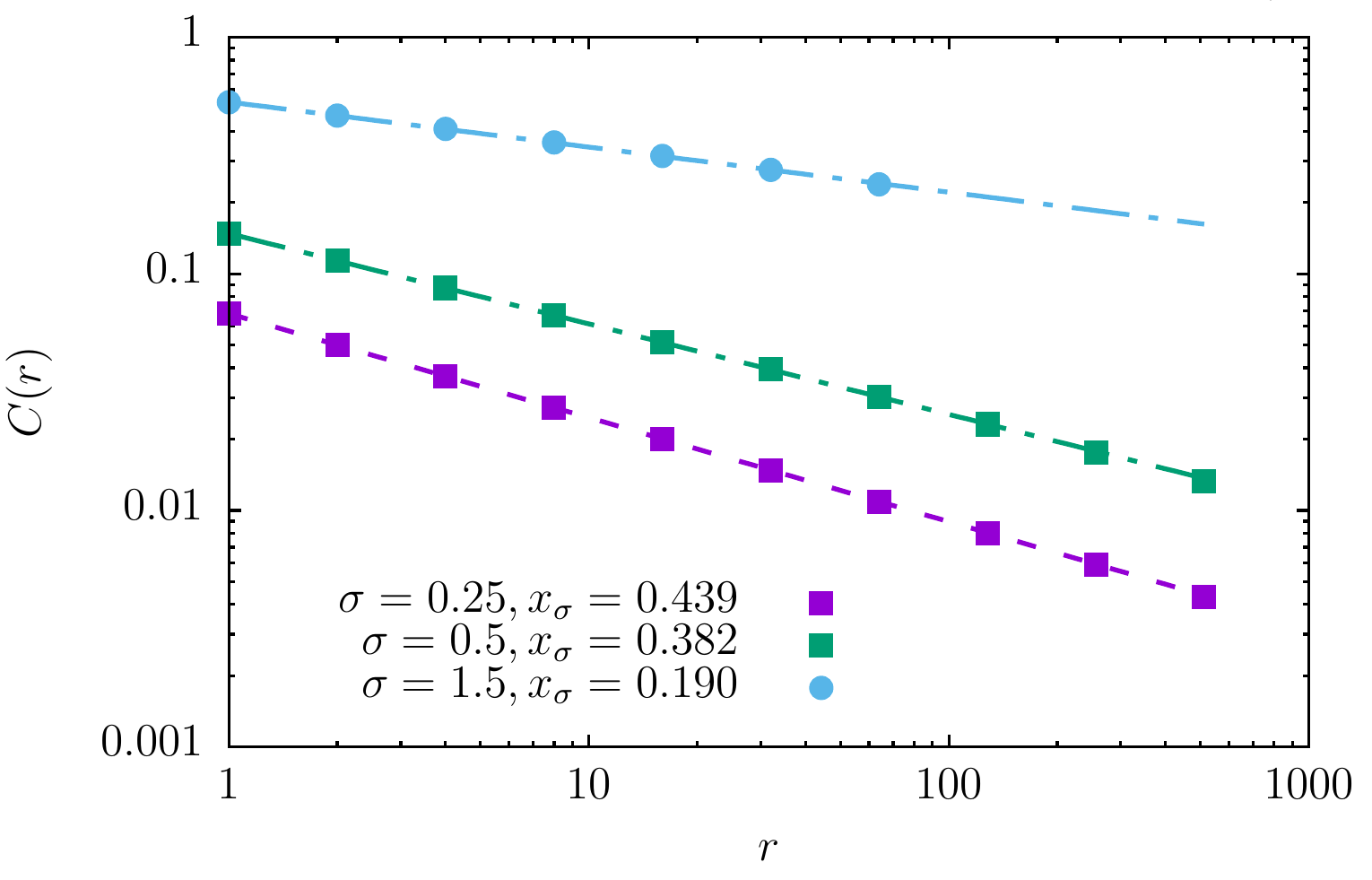}
\caption{Power-law decay of the correlation function for the RG ground state at the critical point. 
$C(r)\sim r^{-2 x_{\sigma}}$ decays with the critical exponent $x_\sigma$ predicted by RG (\ref{xpur}). 
In the plot, we show $C(r)$ as calculated with the recursive reduced density matrix of the partition (\ref{sec:1}) and we compare it with the RG algebraic 
decay (\ref{xpur}). }
\label{fig:correcorre}
\end{figure}

\subsection{Power-law decaying correlation functions.}
\label{sec:corr}

The most striking and surprising aspect of our result is that we have an area law state which supposedly capture power-law correlation functions. 
This fact is against many common beliefs in the literature. 
Such peculiar behaviour is due to the hierarchical structure of the ground state and, a fortiori, to the lack of translational invariance.
Yet, one can be very suspicious whether this is really possible. 
For this reason, from the finite rank reduced density matrix in Eq. (\ref{eq:rho_part1}) we reconstruct the two-point correlation function 
\be
 C(r) \equiv \langle \hat \sigma^x_{\frac{L}{2}+1}\hat \sigma^x_{\frac{L}{2}+1+r} \rangle 
     = \Tr \left [ \hat \rho_{\frac L2} \,\,\hat \sigma^x_{\frac L2+1}\hat \sigma^x_{\frac L2+1+r}
          \right ] \ ,
 \ee
that at the critical point should scale as $C(r) \sim {r^{-2\, x_{\sigma}}}$, with exponent given in Eq. \eqref{xpur}.
We test this behaviour numerically on the fixed-point reduced density matrix obtained recursively.
The $i$-th physical spin is equally correlated with all the physical spins belonging to the same block and
hence, we evaluate the correlation function between effective spins at distance $r=2^k$, being the reduced density matrix (\ref{eq:rho_part1}) written in the basis of the 
$k$-th effective spins. 
The resulting correlation function is reported in Fig. \ref{fig:correcorre} and displays the expected power-law decay with exponent $x_{\sigma}$ predicted analytically
by RG (\ref{xpur}). 

%

\section{Conclusions and perspectives}
In this paper, we presented a detailed RG analysis of the ground-state entanglement entropy of a spin-block in the quantum Dyson hierarchical model. 
Our main goal was to get an analytical insight on the entanglement of long-range interacting spin systems beyond mean-field approximation. 
We found, surprisingly, that entanglement entropy obeys the area law also at criticality when correlation functions decay algebraically. 
This peculiar behaviour is due to the particular simple structure of the RG ground state:  it is a tree tensor network with finite bond dimension and hence it has 
a finite-rank reduced density matrix. 
We must mention that unusual scalings of the entanglement in critical (but non conformal) ground states have been already observed in other models  
\cite{frerot2017entanglement, swingle2016area,aefz-19,cscp-16,ms-16}, 
and so our results represent yet another example of anomalous scaling in the absence of conformal invariance. 
Although it is unlikely that the true long-range ferromagnetic Ising model obeys the area law at criticality,
it is desirable to check this expectation from direct numerical  simulations.

Because of its simplicity, the Dyson hierarchical Hamiltonian is an interesting playground to explore also other entanglement properties of long-range interacting spin chains. 
For example, it would be worthy to characterise better the simple structure of this RG state as a tree tensor network, 
since it undergoes a quantum phase transition with finite entanglement, as those introduced in \cite{wolf2006quantum}. 
It would be interesting also to study the entanglement entropy and negativity between  two disjoint blocks (as in \cite{cct-09,cct-neg} for short range critical systems) 
to investigate whether there is some form of long distance entanglement. 
Finally, the RG approach also gives access to the low-lying excited states and consequently to their entanglement.
In the short-range models, the entanglement of low-lying states captures many interesting physical features (see, e.g., \cite{abs-11}) and it is natural to wonder whether the
same is true for long-range systems.

Finally, questions about the real-time dynamics of long-range systems can in principle be addressed using this model. 
In particular, in mean-field long-range systems, the entanglement entropy after a quench grows logarithmically with time \cite{lerose2018logarithmic}, 
rather than linearly (i.e. ballistically) as for short-range Hamiltonians \cite{cc-05,ac-16,hk-13,nahum-17,bkp-19}.
A number of studies, e.g. \cite{Daley, DaleyEssler}, have reported  numerical evidence of a sub-linear growth also in one spatial dimension. 
Hence, this model appears as the suitable setup to explore the entanglement entropy dynamics with non-equilibrium techniques such as strong 
disorder RG \cite{isl-12,vosk2013many,zas-16,vm-15}.

\ack
We acknowledge useful discussions with Amir Mohammad Aghaei, Vincenzo Alba, Rosario Fazio, Saverio Pascazio, Luca Tagliacozzo, Andrea Trombettoni,  and Xhek Turkeshi.
In particular we thank Luca Tagliacozzo for sharing with us his results \cite{t-prep} before publication. 
We acknowledge support from the European Research Council (ERC) under the EU Horizon 2020 research and innovation program
grant agreements No 694925 (GP) and No 771536 (PC). 
Part of this work has been carried out during the workshop  ``Entanglement in quantum systems'' at the Galileo Galilei Institute (GGI) in Florence.
SP thanks Boston University's Condensed Matter Theory Visitors program for support.

\appendix

\section{Details of calculations}
\label{app:profs_dummies}

In this appendix we report the explicit calculations of some results presented in the main text. 
In \ref{app:prood_powe_two} we derive the recursive structure of the reduced density matrix for the interval $\ell_k=2^{n-k}$ (\ref{eq:alpha_basis}) in partition (\ref{sec:0}). 
In \ref{app:p1proof}, we derive the asymptotic expansions of the entanglement entropy for $\ell=2^m+1$ in Eqs. (\ref{eq:exact_smallsigma}) and (\ref{eq:exact_bigsigma}). 
In \ref{app:finite_redu} we show the finiteness of the rank of the reduced density matrix (cf. Eq. (\ref{eq:rank1})) by linear algebra methods.

\subsection{Recurrence relations for the reduced density matrix in partition (\ref{sec:0}).}
\label{app:prood_powe_two}

Here we derive explicitly the recurrence relations in Eqs. (\ref{eq:alpha_basis}) and (\ref{cdk}). 
We first compute the reduced density matrix of half-system in the ground state $\ket {+} \bra {+}^{[n]}$ (cf.  Eq. (\ref{eq:recu+}))
\begin{equation}
\label{eq:hald_recuded}
\ro_{L/2}\equiv \tr_{B} \ket {+} \bra {+}^{[n]}=\hat \tau_0^{[1]}=   a_{n-1} ^2\,\, \ket{+}\bra{+}_1^{[n-1]} +   b_{n-1} ^2\,\, \ket{-}\bra{-}_1^{[n-1]}\ ,
\end{equation}
where $\hat \tau_0^{[k]}$ is defined in Eq. (\ref{eq:alpha_def}) and $\ket{\pm}_1^{[n-1]}$ is the basis of the first effective spin at level $n-1$. 
The reduced density matrix for $\ell_2=L/4$ is obtained first by rewriting $\hat \tau_0^{[1]}$ in terms of the effective spins at the level $n-2$ as 
\begin{align}
\hat \tau_0^{[1]} & =  c_1  \ket{+}\bra{+}_1^{[n-1]} +
     d_1  \ket{-}\bra{-}_1^{[n-1]} \nn  \\
    & =   c_1  \left (a_{n-2} \ket{+ \,+\,}_1^{[n-2]} +b_{n-1} \ket{- \,-\,}_1^{[n-2]}
    \right ) 
     \left (a_{n-2} \bra{+ \,+}_1^{[n-2]} +b_{n-2} \bra{- \,-}_1^{[n-2]}\,\,
    \right ) \nn \\
    & \quad + \frac 12
     d_1   \left ( \ket{+ \,+\,}_1^{[n-2]} + \ket{- \,-\,}_1^{[n-2]}
    \right ) 
     \left (\bra{+ +}_1^{[n-2]} +\, \bra{- \,-}_1^{[n-2]}\,\,
    \right ) \nn \ ,
\end{align}
and then tracing out the second spin at level $n-2$, 
obtaining
\be
\ro_{L/4}=\hat \tau_0^{[2]}  = \Big ( c_1   a_{n-2} ^2 + \frac 12  d_1  \Big)
    \ket{+}\bra{+}_1^{[n-2]} +
     \Big ( c_1   b_{n-2}^2  + \frac 12  d_1 \Big)
    \ket{-}\bra{-}_1^{[n-2]} \ ,
\ee   
which is Eq. (\ref{eq:alpha_basis}) with $k=2$. Iterating this procedure, one easily gets the recurrence relation for $\ro_{L/2^k}$.
At the critical point the coefficients in Eq. (\ref{eq:state_coeff}) are constant for all $k$. All the recursion relations simplify and admit as solution Eq. (\ref{eq:alpha_solution}).

\subsection{Asymptotic expansions of the entanglement entropy for partition (\ref{Sec:2})}
\label{app:p1proof}
Here we work out the asymptotic expressions of the entanglement entropy for $\ell_m=2^m+1$ in Eqs. (\ref{eq:exact_smallsigma}-\ref{eq:exact_bigsigma}) 
in the limits $\sigma \gg1$ and $\sigma \ll1$.
In this case the reduced density matrix (\ref{eq:rho_pm}) for $p=1$, i.e.
\begin{equation}
\hat \rho_{\ell_{m}} = c_{\nu}^+ {\ro}^{[1,m]}_0  + d_{\nu}^+ \ro^{[1,m]}_1,
\end{equation}
is a $4\times 4$ block-diagonal matrix in the basis $ \ket{\pm}^{[m]}_1 \otimes \ket {\pm}^{[0]}_{\ell_m}$, which makes all calculations straightforward.  
This equation can be solved using the recurrence relations (\ref{eq:properties_wowo_3}) with the initial conditions (\ref{eq:alpha_solution}), obtaining 
\begin{equation} \label{eq:rho3}
\hat \rho_{\ell_{m}} = 
\begin{pmatrix}
    \ro^s_{\nu, m} & 0\\
    0 &    \ro^a_{\nu, m}
\end{pmatrix} \ ,
\end{equation}
with $\ro^{s/a}_{\nu, m}$ block matrices of the form
\begin{equation}\label{eccola}
\ro^{s/a}_{\nu,m} =
\begin{pmatrix}
A_{\nu, m}^{s/a} & C_{\nu, m}^{s/a} \\
C_{\nu, m}^{s/a} & B_{\nu, m}^{s/a}
\end{pmatrix}  =
\begin{pmatrix}
A_{\nu}^{s/a} + D_{\nu}^{s/a} e^{-\alpha m}& C^+_{\nu} e^{-\beta m} \pm C^{-}_{\nu}e^{-\gamma m} \\
C^+_{\nu} e^{-\beta m} \pm C^{-}_{\nu}e^{-\gamma m} & B_{\nu}^{s/a} + E_{\nu}^{s/a} e^{-\alpha m}
\end{pmatrix} \ ,
\end{equation}
where  $A^{s/a}_{\nu}, \cdots E_{\nu}^{s/a}$ are read from Eqs. (\ref{eq:alpha_solution}) and (\ref{eq:rho_pm}). 
We take now the limit $n \to \infty$ and, since $\nu = n-p-m$, the elements $ c^+_{\nu} ,\, d^+_{\nu} $ are constant and given by the limit for $k\to\infty$ of Eq. (\ref{eq:alpha_solution}). 
Defining $\xi \equiv \sqrt{K^2+4}$, the coefficients in (\ref{eccola}) are 
\begin{subequations}\label{eq:coeffi_xi}
\begin{align}
A^s & = \frac{\xi^2}{4 (\xi-1)^2} \ , \quad\quad
A^a = B^a=\frac{\xi^2(1+K^2)-2\xi}{4(\xi^2-1)^2} \ , \qquad
B^s  =  \frac{(\xi-2)^2}{4(\xi-1)^2} \ ,
\\
D^s & =E^s= \frac{K^2}{4 (\xi^2(1+K^2) + 2 \xi )}      \ , \quad \quad
D^a =E^a= -D^s \ , \\
C^{\pm} & = \frac{
\sqrt{\xi^2-2\xi}
 \pm \sqrt{\xi^2+2 \xi} \mp 2 \sqrt{1+2\xi}
 }{8 (\xi-1)} \ .
\end{align}
\end{subequations}
The reduced density matrix can be diagonalised  and from its four eigenvalues $\{\lambda_i\}$, the entanglement entropy 
$S_{A}  = -\sum_{i=1}^4 \lambda_i \ln \lambda_i$ is obtained as a function  of $K$. 
In the limits $\sigma\gg1$ and $\sigma \ll 1$, we derive the asymptotic expression of $S_{\ell_{m}}$ in Eqs.~(\ref{eq:exact_smallsigma}-\ref{eq:exact_bigsigma}).

{\it Asymptotic result for $\sigma\ll1$.}
We consider the limit $\sigma \ll 1$, equivalent to $K\to 0$, see Eq. (\ref{eq:flow_rg}). 
At order $\mathcal O(K^3)$ the matrix elements (\ref{eccola}) read 
\begin{subequations}\label{eq:777}
\begin{align}
A_m^s & = 1 -\frac{K^2}4 + \frac{K^2}{32}\frac 1{2^m} + \mathcal O(K^{3})   , \qquad
B_m^s  = \frac{K^2}{32}\frac1{2^m} + \mathcal O(K^{4})  ,\\
C_m^s & = \frac{K}{{16 \sqrt 2}}\, \frac1{ 2^{m/2}}+ \mathcal O(K^{3})  , \qquad
C_m^a  =\frac{K^2}{{32 \sqrt 2}} \frac{(m+2)}{2^{m/2}}+ \mathcal O(K^{3})  .\label{66:6}
 \\ 
 A_m^a &=B_m^a = \frac{K^2}8 - \frac{K^2}{32}\frac 1{2^m} + \mathcal O(K^{3})  .\qquad
    \end{align}
\end{subequations}
With these elements, the eigenvalues of the symmetric and anti-symmetric matrices (\ref{eccola})  are
\begin{align}
& \lambda_1  =  \mathcal O(K^{3}) \ , \\
& \lambda_2  = 1-\frac{K^2}4 \left ( 1-2^{-(m+2)} \right ) + \mathcal O(K^{3})\ ,\nn
\\
& \lambda_{3,4}  = \frac{K^2}{8} \left( 1 - 2^{-(m+2)} \pm  2^{(m+3)/2}(m+2)
\right) + \mathcal O(K^{3})  ,\nn
\end{align}
hence
the asymptotic entanglement entropy as a function of $m$ reads
\begin{equation}
S_{\ell_{m}} = \frac {K^2}{4} \Big
    (1-\ln \frac{K^2}8 \Big) 
    +\frac {K^2}{16}\frac1{ 2^m}
      \Big (\ln \frac {K^2}{8} -\frac{(m+2)^2}4 
      \Big) + \mathcal O(K^3) \ ,
 \end{equation}
that, as function of $2^m = \ell_{m} -1 $, is the same as Eq. (\ref{eq:exact_smallsigma}).

{\it Asymptotic result for $\sigma\gg1$.}
Here we consider the limit $\sigma \gg 1$, i.e. for  $K\to \infty$, and we expand the matrix elements (\ref{eccola}) at $\mathcal O(K^{-3})$. 
The saturation value is obtained for $\alpha\, m ,\beta m, \gamma \,m \gg 1$, corresponding to have only diagonal terms in Eq. (\ref{eccola}) given by
\begin{subequations}\label{eq:ppp}
\begin{align}
A^s & =\frac 14 + \frac 1{2K} + \frac 3{4 K^2}  + \mathcal O(K^{-3}) \ ,
 \\ 
B^s & =\frac 14 - \frac 1{2K} - \frac 1{4 K^2} + \mathcal O(K^{-3})\ ,
 \\
 A^a &=B^a =\frac 14 - \frac 1{4 K^2}  + \mathcal O(K^{-3}) \ ,
\end{align}
\end{subequations}
leading to the following saturation entanglement entropy
\begin{equation}
S_{\text{sat}} = 2 \ln 2 -\frac 1{K^2} + \mathcal O(K^{-3}) \ .
\end{equation}

The preasymptotic regime, is obtained by taking the limit $K\to\infty$ at fixed $m$. 
In this limit 
\begin{equation}
\alpha = 2\ln K + \frac 4{K^2} + \mathcal O(K^{-3}), \quad \quad 
\beta = \frac 1{2K^2} + \mathcal O(K^{-3}), \quad \quad
\gamma=\ln K + \frac 3{2 K^2} + \mathcal O(K^{-3}) \\ ,
\end{equation}
and the exponent $\beta \to 0$, so that we cannot neglect  the non-diagonal parts of Eq. (\ref{eccola}).
Thus, there is another regime in which $m\gg \alpha^{-1}, \gamma^{-1}$ but with  $\beta m\ll 1$, i.e., in terms of $K$,  $(\ln K)^{-1}\ll m \ll 2 K^2$. 
In this regime, the diagonal terms $A_m^{s/a}, B_m^{s/a}$ are the same as in Eqs. (\ref{eq:ppp}), but the non-diagonal ones become
\begin{align}
C_m^s &=C_m^a =\left(\frac 14 - \frac {3+m}{8K^2} \right) +\mathcal O(K^{-3}, m^2K^{-4}).  
    \end{align}
The diagonalisation of  the reduced density matrix (\ref{eq:rho3}) finally  yields
\begin{equation}
S_A  = \ln 2 +\frac{m}{4K^2} \Big ( 1 -  \ln \frac{m}{4 K^2}  \Big ) + \mathcal O\left (K^{-3}, m^2K^{-4}, K^{-m}\right)  , 
\end{equation}
that in terms of $2^m = \ell_{m} -1 $ is Eq. (\ref{eq:exact_bigsigma}) in the main text.

\subsection{Rank of reduced density matrices.}
\label{app:finite_redu}

Here we show Eq. (\ref{eq:rank1}) of the main text. All the matrices $\ro_i^{[n]}$ in Eq. (\ref{eq:recurrence_matrix}) can be written in a compact form as 
\begin{equation}
\label{eq:rho1_rec}
\hat \rho_i^{[n]} = \sum_{j,k=0}^3 \gamma_{jk}^i \hat P_j \otimes \, \hat \rho_k^{[n]} \ ,
\end{equation}
where 
\begin{equation}
\label{eq:properties_wowo_1}
 \hat P_0= \sigmaup \ , \quad \hat P_1=\sigmadown\ , \quad\hat P_2=\sigmaplus \ ,\quad \hat P_3 = \hat P_2^T ,
\end{equation}
while the real coefficients of the tensor $\gamma^i_{jk}$ can be obtained from Eq. (\ref{eq:recurrence_matrix}). For fixed $i,j$ there is only one $k_{ij}$ such that $\gamma^i_{jk_{ij}}\neq 0$ (e.g. $\gamma^0_{0k} = a^2 \, \delta_{k0}$).  The coefficients $\gamma^i_{jk_{ij}}$ satisfy the following property
\begin{equation}
\label{eq:properties_wowo_2}
\quad \gamma_{0,k_{0i}}^i\,\gamma_{1,k_{1i}}^i- \gamma_{2,k_{2i}}^i\,  \gamma_{3,k_{3i}}^i =0  \quad \forall i \ .
\end{equation} 
Eq. (\ref{eq:rho1_rec}) is written in the basis of the spin blocks at level $n$, see Fig. \ref{fig:tree_scaling} 
(e.g. $\hat\rho_i^{[n=0]}$ is written in terms of the physical spins). 
For generic $n$, $\hat \rho_i^{[n]}$ is a matrix of size $2^{n+1}\times 2^{n+1}$.
By construction, up to a change of basis, the matrices $\rho_i^{[n]}$ are either diagonal or anti-diagonal 
\begin{equation}
\rho_{0,1}^{[n]} = {
    \begin{pmatrix}
    A^{[n]}_{0,1} & 0\\
    0 & B^{[n]}_{0,1} 
    \end{pmatrix}}
    \ ,\quad 
\rho_{2}^{[n]} = {
    \begin{pmatrix}
    0 & C_2^{[n]}\\
    D_2^{[n]} & 0 
    \end{pmatrix}}
\quad  \text{and}\quad 
\rho_3^{[n]} = \big(\rho_2^{[n]}\big)^T .
\end{equation}
In order to prove Eq. (\ref{eq:rank1}), we show that $\forall n$ it holds 
\begin{subequations} \label{eq:pippopalla}
\begin{align}
& A_0^{[n]}\, B_1^{[n]} =  C_2^{[n]}D_3^{[n]}\quad \text{and} \quad A_1^{[n]}\, B_0^{[n]} = \left ( D_2^{[n]}\right )^2 \ , \\
& \Rank\left[ A_{0,1}^{[n]}\right ]= \Rank\left[B_{0,1}^{[n]}\right]=\Rank\left[C_{2}^{[n]}\right]=\Rank\left[D_{2}^{[n]}\right]=1 \ , \quad
\Rank\left[\rho_i^{[n]}\right]=2 \ .
\end{align}
\end{subequations}

\begin{proof} We will prove by induction. The step $n=0$ is obvious.  Let us first see what happens for $n=1$ in order to start the induction. 
We construct the matrices $\ro_i^{[1]}$ with $\gamma^i_{jk}$ given by Eq. (\ref{eq:recurrence_matrix}). Let us consider
\begin{align}
\rho_{0}^{[1]} = {
    \begin{pmatrix}
    \gamma^0_{0, k_{00}} A^{[0]}_{0} & 0 & 0 &\gamma^0_{2, k_{02}} C_2^{[0]}\\
    0  & \gamma^0_{0, k_{00}} B^{[0]}_{0}  & \gamma^0_{2, k_{02}} D_2^{[0]} & 0\\
    0  & \gamma^0_{3, k_{30}} C^{[0]}_{3}  & \gamma^0_{1, k_{01}} A_1^{[0]} & 0\\
    \gamma^0_{3, k_{30}}  D^{[0]}_3 & 0 & 0 &\gamma^0_{1, k_{01}} B_1^{[0]}\\    
    \end{pmatrix}}
\end{align}
and analogously for the other $\rho_i^{[1]}$. After a change of basis, this defines the recurrence for the matrices $A, B, C, D$ as
\begin{align}
\label{eq:recursive_A}
&A_0^{[1]} = 
\begin{pmatrix}
 \gamma^0_{0, k_{00}} A^{[0]}_{0} & \gamma^0_{2, k_{02}} C_2^{[0]} \\
 \gamma^0_{3, k_{03}}  D^{[0]}_3 &  \gamma^0_{1, k_{01}} B_1^{[0]} 
\end{pmatrix}
 \qquad 
B_0^{[1]} = 
\begin{pmatrix}
 \gamma^0_{1, k_{01}} A^{[0]}_{1} & \gamma^0_{3, k_{03}} C_3^{[0]} \\
 \gamma^0_{2, k_{02}}  D^{[0]}_2 &  \gamma^0_{0, k_{00}} B_0^{[0]} 
\end{pmatrix}   \\
\quad 
&A_1^{[1]} = 
\begin{pmatrix}
 \gamma^1_{0, k_{10}} A^{[0]}_{1} & \gamma^1_{2, k_{12}} C_3^{[0]} \\
 \gamma^1_{3, k_{13}}  D^{[0]}_2 &  \gamma^1_{1, k_{11}} B_0^{[0]} 
\end{pmatrix}
 \qquad 
B_1^{[1]} = 
\begin{pmatrix}
 \gamma^1_{0, k_{11}} A^{[0]}_{0} & \gamma^1_{3, k_{13}} C_2^{[0]} \\
 \gamma^1_{2, k_{12}}  D^{[0]}_3 &  \gamma^1_{0, k_{10}} B_1^{[0]} 
\end{pmatrix} \nn \\
& C_2^{[1]} = 
\begin{pmatrix}
 \gamma^2_{0, k_{20}} A^{[0]}_{0} & \gamma^2_{2, k_{23}} C_2^{[0]} \\
 \gamma^2_{3, k_{23}}  D^{[0]}_3 &  \gamma^2_{1, k_{21}} B_1^{[0]} 
\end{pmatrix}
 \qquad 
D_2^{[1]} = 
\begin{pmatrix}
 \gamma^2_{1, k_{21}} A^{[0]}_{1} & \gamma^2_{3, k_{23}} C_3^{[0]} \\
 \gamma^2_{2, k_{22}}  D^{[0]}_2 &  \gamma^2_{0, k_{20}} B_0^{[0]} 
\end{pmatrix} \nn
\end{align}
These matrices have rank one if the determinant is zero (since they are not zero, the rank is not null). 
For the matrix $A_0^{[1]}$ we have
\begin{multline}
\det A^{[1]}_0  =  
\gamma^0_{1, k_{01}}  \gamma^0_{1, k_{01}} A^{[0]}_{1} B_1^{[0]} - 
\gamma^0_{2, k_{02}} \gamma^0_{3, k_{03}}  C_2^{[0]} D^{[0]}_3 
 = \\=
A^{[0]}_{1} B_1^{[0]} \left (\gamma^0_{1, k_{01}}  \gamma^0_{1, k_{01}} -
\gamma^0_{2, k_{02}} \gamma^0_{3, k_{03}}  \right ) = 0\ .
\end{multline}
Because of Eqs. (\ref{eq:properties_wowo_3}-\ref{eq:properties_wowo_1}-\ref{eq:properties_wowo_2}), the same is true for all the matrices.

Let us now assume Eq. (\ref{eq:pippopalla}) to be true at $n-1$ and show that the same holds at level $n$. 
Eq. (\ref{eq:recursive_A}) holds at every step after a permutation of the basis elements. We now want to prove that they have rank one. Let us re-write the block matrices $A_0^{[n]}$ setting $ \gamma^0_{0, k_{00}} / \gamma^0_{2, k_{02}} = \gamma^0_{3, k_{03}} /  \gamma^0_{1, k_{01}} = w_0$. Up to a multiplication of the first and the second rows by a coefficient, we have
\be
A_0^{[n]} = 
\begin{pmatrix}
   w_0 A^{[n-1]}_{0} &  C_2^{[n-1]}  \\
   w_0 D^{[n-1]}_3 & B_1^{[n-1]} 
\end{pmatrix} \to 
\begin{pmatrix}
   w_0 A^{[n-1]}_{0} B_1^{[n-1]} &  B_1^{[n-1]} C_2^{[n-1]}  \\
   w_0 C_2^{[n-1]} D^{[n-1]}_3 & C_2^{[n-1]} B_1^{[n-1]} 
\end{pmatrix} \ ,
\ee
where we multiplied the first row by the block matrix $B_1^{[n-1]}$ on the right and of the second row by the block matrix $C_2^{[n-1]}$ on the left. Since we assumed  
$A_0^{[m]}\, B_1^{[m]} =  C_2^{[m]}D_3^{[m]}$ to hold at $n-1$, this means that the second row is linearly dependent on the first, hence the rank is the one of the first rows $\Rank [A_0^{[n]}] = \Rank [\left (w_0 A_0^{[n-1]},B_1^{[n-1]} \right ) ] = \min \left (\Rank[A_0^{[n-1]}],\Rank[B_1^{[n-1]}\right) = 1 $. It is immediate to verify that ${A_0^{[n]}\, B_1^{[n]} =  C_2^{[n]}D_3^{[n]}}$.

\end{proof}


\end{document}